\documentclass[lettersize,journal]{IEEEtran}
\usepackage{amsmath,amsfonts}

\usepackage{algorithmic}
\usepackage{algorithm}
\usepackage{array}
\usepackage{textcomp}
\usepackage{stfloats}
\usepackage{url}
\usepackage{verbatim}
\usepackage{graphicx}
\usepackage{cite}
\usepackage{xcolor}
\usepackage{subfigure}
\usepackage{booktabs}
\usepackage{multirow}
\begin{document}

\title{Deep Learning-based Human Gesture Channel Modeling for Integrated Sensing and Communication Scenarios}

\author{Zhengyu~Zhang,~\IEEEmembership{Student Member,~IEEE,}
        Neeraj~Varshney,~\IEEEmembership{}
        Jelena~Senic,~\IEEEmembership{}\\
        Raied~Caromi,~\IEEEmembership{}
        Samuel~Berweger,~\IEEEmembership{}
        Camillo~Gentile,~\IEEEmembership{Member,~IEEE,}
        Enrico~M.~Vitucci,~\IEEEmembership{Senior Member,~IEEE,}
        Ruisi~He,~\IEEEmembership{Senior Member,~IEEE,}
        Vittorio~Degli-Esposti,~\IEEEmembership{Senior Member,~IEEE}
        
\thanks{

Zhengyu Zhang, Enrico M. Vitucci and Vittorio Degli-Esposti are with the Department of Electrical, Electronic, and Information Engineering “Guglielmo Marconi” (DEI), CNIT, University of Bologna, 40136 Bologna, Italy. (e-mail: \{zhengyu.zhang; enricomaria.vitucci; v.degliesposti\}@unibo.it)

N. Varshney, R. Caromi, and C. Gentile are with the Communications Technology Laboratory, National Institute of Standards and Technology (NIST), Gaithersburg, MD 20899. N. Varshney is also with Prometheus Computing LLC, Cullowhee, NC 28723. J. Senic and S. Berweger are with the Communications Technology Laboratory, NIST, Boulder, CO 80305. (e-mail: \{neeraj.varshney; jelena.senic; raied.caromi; samuel.berweger; camillo.gentile\}@nist.gov)

Ruisi He is with the School of Electronics and Information Engineering, Beijing Jiaotong University, Beijing 100044, China. (e-mail: ruisi.he@bjtu.edu.cn)
}}

\markboth{Journal of \LaTeX\ Class Files,~Vol.~14, No.~8, August~2021}%
{Shell \MakeLowercase{\textit{et al.}}: A Sample Article Using IEEEtran.cls for IEEE Journals}


\maketitle

\begin{abstract}
With the development of Integrated Sensing and Communication (ISAC) for Sixth-Generation (6G) wireless systems, contactless human recognition has emerged as one of the key application scenarios. Since human gesture motion induces subtle and random variations in wireless multipath propagation, how to accurately model human gesture channels has become a crucial issue for the design and validation of ISAC systems. To this end, this paper proposes a deep learning-based human gesture channel modeling framework for ISAC scenarios, in which the human body is decomposed into multiple body parts, and the mapping between human gestures and their corresponding multipath characteristics is learned from real-world measurements. Specifically, a Poisson neural network is employed to predict the number of Multi-Path Components (MPCs) for each human body part, while Conditional Variational Auto-Encoders (C-VAEs) are reused to generate the scattering points, which are further used to reconstruct continuous channel impulse responses and micro-Doppler signatures. Simulation results demonstrate that the proposed method achieves high accuracy and generalization across different gestures and subjects, providing an interpretable approach for data augmentation and the evaluation of gesture-based ISAC systems.

\end{abstract}


\section{Introduction}

\IEEEPARstart{W}{ith} the development of Sixth-Generation (6G) mobile communications, wireless systems are progressing toward higher frequency bands, broader bandwidths, and more intelligent integration, enabling ubiquitous connectivity, sensing, and intelligence\cite{Ref1}. Among the emerging technologies, Integrated Sensing and Communication (ISAC) has attracted significant attention, as it allows wireless signals to simultaneously support both communication and sensing tasks, thereby enhancing spectral efficiency and environmental awareness\cite{Ref2,Ref3,Ref.add3}. As part of the interest in ISAC, contactless human gesture recognition has emerged as a representative application, attracting increasing interest from both academia and industry. By leveraging Machine Learning (ML) to analyze and model complex motion patterns, systems can perceive and understand diverse human activities, enabling promising applications in smart interaction, health monitoring, and other emerging 6G scenarios\cite{Ref4,Ref5}. 

In ISAC scenarios, the wireless channel not only serves as the transmission medium\cite{Ref.add4} but also functions as the information source for environmental sensing\cite{Ref6}. Particularly in gesture recognition, body movement causes significant variations in the propagation environment, giving rise to distinct multipath signatures that are tightly coupled with human dynamics\cite{Ref7,Ref8,Ref9}. Accurate modeling of such gesture-based channels is thus fundamental for reliable sensing, robust gesture recognition, and system design\cite{Ref10}.

On the other hand, ML-based gesture recognition has made impressive progress by leveraging large-scale and diverse datasets. However, collecting real-world data is expensive, time-consuming, and often constrained by privacy concerns\cite{Ref11}. Therefore, developing interpretable and generalizable channel models would not only provide theoretical insights but would also serve as a powerful tool for data augmentation, enhancing ML robustness under a variety of gestures.

Nevertheless, existing human gesture channel modeling methods remain limited. Statistical models describe channel responses through averaged characteristics, failing to capture detailed variations caused by gestures\cite{Ref12}. Deterministic approaches, such as ray-tracing, are often based on theoretical propagation formulas and struggle to handle the actual complexity,  randomness and non-geometric nature of the human body\cite{Ref13}. Furthermore, human gestures are highly diverse and critical, even subtle movements can significantly affect the gesture channels\cite{Ref14}. Unfortunately, most existing models fall short in interpretability and generalization, making them difficult to apply across diverse subjects and time-varying behaviors.

Recent studies in wireless human sensing have explored features such as Received Signal Strength (RSS), Channel State Information (CSI), and Doppler shifts to classify activities. Abdelnasser et al. \cite{Ref18} used Wi-Fi RSS for hand gesture recognition, while Ali et al. \cite{Ref19} detected keystrokes via CSI variations caused by finger micro-movements. M. Li et al. \cite{Ref20} exploited range-velocity-time signatures for behavioral understanding. Micro-Doppler features were used in Reference \cite{Ref21} for non-contact breathing monitoring and in Reference \cite{Ref22} for classifying subtle human activities. However, most studies focus on behavior sensing rather than reconstructing channel responses, often relying on statistical or heuristic methods without generalizable models, limiting their applicability to diverse real-world scenarios. Several works have explored micro-Doppler modeling, which serves as the core dynamic characteristic of human gesture channels. Fairchild et al. \cite{Ref24} analytically synthesized Doppler shifts from human motion trajectories, while Abdelgawwad et al. \cite{Ref25} proposed a 3D deterministic model using point scatterers for body segments. Liu et al. \cite{Ref26} combined trajectory-driven motion with a stochastic scattering environment. Shi et al. \cite{Ref27} simulated micro-Doppler via electromagnetic scattering, and Erol et al. \cite{Ref28} generated radar echoes using the sensor "Kinect" by Microsoft. However, these studies rely on simplified, fixed geometric models, resulting in synthesized signatures that are a bit stilted and not very natural. Meanwhile, they overlook the joint nature of other critical channel characteristics—such as delay, angle, and path gain, which could affect the overall consistency of human gesture channel modeling.

To overcome these problems, we need a data-driven approach that can capture the fine-grained dynamics of human motion while maintaining the physical coherence across multiple channel dimensions. Deep learning (DL) provides a promising solution by learning high-dimensional, nonlinear mappings between environmental characteristics and channel responses, making it a powerful tool for capturing complex multipath propagation effects\cite{Ref15,Ref16,Ref17}. Zhang et al. \cite{Ref29} proposed a Generative Adversarial Network (GAN)-based digital twin to generate Channel Impulse Response (CIR) matching statistical properties like path loss and delay spread. Xiao et al. \cite{Ref30} learned delay and antenna-domain power distributions for Multiple-Input Multiple-Output (MIMO) CIR synthesis. In other words, generative networks have also been applied to human gesture channel modeling. Rahman et al. \cite{Ref31} used GANs to generate micro-Doppler spectrograms, while Alnujaim et al. \cite{Ref32} employed Pix2Pix, a conditional GAN which can learn how to map input images to output images, to synthesize spectrograms across different aspect angles. However, these works focus on directly generating characteristics, and lack the capacity to model multipath-level channels. Besides, data-driven methods often suffer from limited interpretability, making channel models difficult to generalize across sets of diverse human gestures.

Motivated by these challenges, this paper proposes a unified deep learning-based framework for gesture-based channel modeling in ISAC scenarios. First, the human body is decomposed into six major body parts, and the multipath distribution of each is independently learned under the assumption of single-bounce scattering. Next, a context-aware channel sounding system at 28 GHz is employed to jointly capture temporally and spatially synchronized signals, RGB images, and Lidar point clouds from human subjects under sixteen gesture conditions. Then, based on the collected real-world measurements, the Space-Alternating Generalized Expectation-maximization (SAGE) algorithm is used to estimate multipath parameters, while computer vision techniques extract human keypoints from the image and point cloud data. A Poisson neural network predicts the number of MPCs, also referred to as \textit{paths}, associated with each of the six major human body parts, and a Conditional Variational Auto-Encoder (C-VAE) generates scattering points to reconstruct continuous, time-varying channels. Experimental results demonstrate the proposed framework's ability to accurately reproduce channel impulse responses and micro-Doppler signatures, achieving strong generalization across different gestures and subjects.

The remainder of this paper is organized as follows. Section II describes the sensing channel measurement setup and campaign. In Section III, the data processing is detailed. Section IV introduces the overall framework of the proposed human gesture channel modeling approach. Section V presents network training and performance results. Section VI provides extensive evaluation results on channel simulation. Section VII concludes the paper.

\section{Sensing Channel Measurements}
\label{sec:sensing}

\begin{figure}
    \centering
    \includegraphics[width=0.5\textwidth]{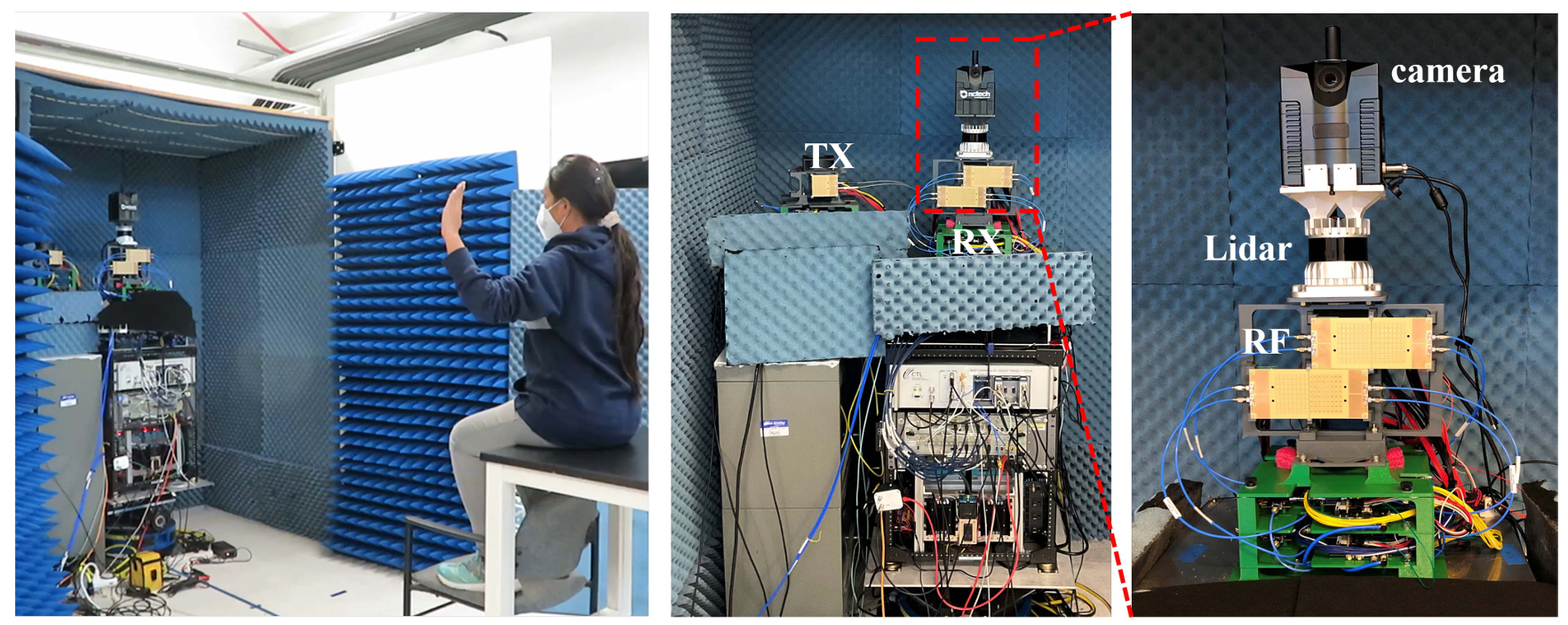} 
    \caption{Measurement setup for gesture-based sensing: The 28~GHz RX includes four phased-array boards arranged in a staggered 2$\times$2 configuration. A panoramic camera and Lidar scanner are mounted above the RX for synchronized visual data capture. All subsystems are temporally and spatially aligned to enable 3D keypoint generation and gesture-based channel modeling.}
    \label{fig1}
\end{figure}

The sensing channel measurements were collected using the context-aware channel sounder described in~\cite{gentile2023context}, with architectural enhancements supporting gesture recognition detailed in~\cite{caromi2024gesture}. The context-aware system combines a 28~GHz RF channel sounder, panoramic camera, and 3D Lidar scanner, all temporally synchronized and spatially aligned to enable MPC extraction and gesture labeling within a shared coordinate system. Figure~\ref{fig1} shows the sounder, with the optical subsystems mounted above the RX phased-array antennas of the RF subsystem.

\subsection{Measurement Equipment}

The RF subsystem uses 8$\times$8 microstrip patch arrays at both transmitter (TX) and receiver (RX), operating at 28.5~GHz with 2~GHz bandwidth. At the RX, four phased-array boards are stacked on-grid to form a 256-element aperture, for enhanced resolution in both azimuth and elevation. During each acquisition, a pseudorandom (PN) sequence is transmitted with a 2.6~ms sweep duration and received as a phase-synchronized CIR across all elements. Beamforming is performed in post-processing to generate a 3D power distribution over azimuth, elevation, and delay, which defines the RF domain used throughout this paper. This power cube is then passed to the SAGE super-resolution algorithm, which extracts discrete MPCs by estimating path gains, delays, and angles-of-arrival \cite{Ref34}. SAGE provides approximately fivefold resolution improvement over the system’s inherent beamwidth and bandwidth, and de-embeds the antenna and front-end responses to isolate the propagation channel. In controlled validation experiments, the average absolute error was 0.2~dB in path gain, 0.1~ns in delay, and 0.2$^\circ$ in both azimuth and elevation~\cite{Ref10}. 

The optical subsystem includes an OS0-128 Lidar and an iSTAR Pulsar panoramic camera.%
\footnote{Certain commercial equipment, instruments, or materials are identified in this paper to specify the experimental procedure adequately. Such identification is not intended to imply recommendation or endorsement by the National Institute of Standards and Technology, nor is it intended to imply that the materials or equipment identified are necessarily the best available for the purpose.} The Lidar captures point clouds at 10~Hz over a 360$^\circ \times$ 90$^\circ$ Field-of-View (FoV), while the camera captures RGB frames at 7~Hz with 11000$\times$5500 resolution. Both systems are mounted above the RF subsystem at the RX to minimize parallax and are calibrated using fiducial markers visible in all modalities. Keypoints extracted from the fused camera and Lidar data are used to label gesture segments and are described in detail in Section~\ref{sec:keypoints}.

Although the three subsystems operate at different acquisition rates, they are temporally synchronized to a shared RF reference clock, allowing all systems to observe and record the same scene simultaneously. This temporal alignment ensures that the RF, Lidar, and camera data are synchronized for joint processing. Additionally, the subsystems are spatially registered in a common 3D geometric coordinate system defined with respect to the RF array aperture. Calibration markers and fixed mounting geometry are used to project RF MPCs, Lidar point clouds, and image-derived keypoints into this shared spatial domain using consistent units of azimuth, elevation, and range, which we obtain directly from the delay. This alignment enables direct fusion of modalities for gesture-based channel modeling.

\subsection{Measurement Campaign}

The dataset consists of synchronized RF and pose measurements from four seated participants (two male, two female), each facing the TX/RX arrays at a distance of 2.5~m, with antenna height set at shoulder level (154~cm). The system, shown in Fig.1, was deployed in a semi-anechoic environment with scatterer clearance to support delay and angular isolation of human-induced MPCs.

The 16 dynamic hand gestures are considered in this study. These involve simultaneous but independent motion of the left and right hands from a resting position (arms extended forward at shoulder level), moving in the four cardinal directions: \textit{up, down, left,} and \textit{right}. Each participant performed all 16 gestures, resulting in 64 labeled gesture sequences in total. Each sequence spans 3.9~s and includes 1500~RF snapshots (27~RGB frames and 39~Lidar frames). These measurements are used to train and evaluate the modeling and classification techniques introduced in Sections~\ref{sec:keypoints} and~\ref{sec:classification}.

\section{Data Pre-processing}
\label{sec:keypoints}

\subsection{Human Gesture Characterization}

To accurately model human gestures, we extract human keypoints models from RGB images. Specifically, the video frames captured by the camera are processed using the High-Resolution Network (HRNet) \cite{Ref33}, which detects 19 keypoints corresponding to critical parts of the human body, including the nose, left eye, right eye, left ear, right ear, left shoulder, right shoulder, left elbow, right elbow, left wrist, right wrist, left hip, right hip, left knee, right knee, left ankle, right ankle, chest and belly.

After extracting the keypoints, we further map them from the pixel coordinates to a three-dimensional coordinate system. In particular, the pixel coordinates in the image are first scaled according to the FoV of the camera to obtain angular information; the resulting azimuth and elevation values are then aligned with the depth maps captured by the Lidar to obtain spatial range. In this way, we achieve the transformation from 2D RGB images to a 3D key-point model. Compared with the channel sounder, the human keypoints are captured at a lower frame rate (determined by the camera). Therefore, we interpolate the keypoint trajectories to align them with the higher RF sampling rate.

To eliminate variations in body size and position across different subjects, we align all keypoints by translating them with respect to a fixed reference point (belly keypoint). This allows us to compare poses across participants in a consistent manner.

Furthermore, to establish a link between MPCs and the physical structure of the human body, we group the nineteen keypoints into six human body parts: Left/Right Forearm (L/R), Left/Right Upper Arm (L/R), Head, and Torso, as illustrated in Figure 2 (In the following, we use “L” and “R” to denote “Left” and “Right,” respectively). We omit the knee and ankle keypoints because they remain stationary during all the gestures. This association relies on the assumption of single-bounce scattering, which allows us to reasonably assign each multipath component to a specific body part. And the grouping provides physical interpretability for the subsequent clustering of scattering points.

\begin{figure}
\centering
\subfigure[]{\includegraphics[width=1.1in]{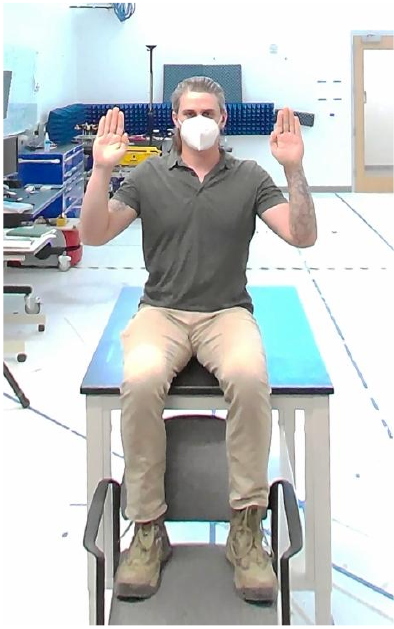}}
\subfigure[]{\includegraphics[width=1.8in]{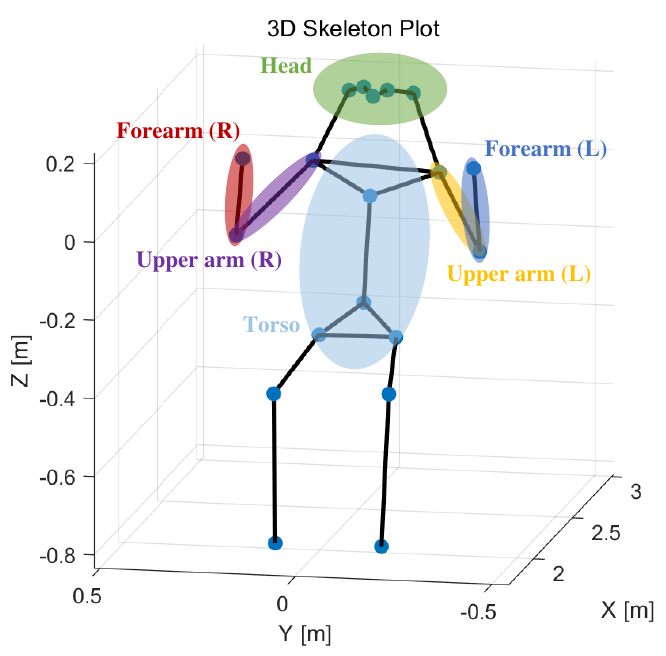}}
\caption{An illustration of the keypoints model. (a) Human gesture. (b) Keypoints model.}
\label{fig2}
\end{figure}

\subsection{Scattering Points Estimation}
\label{sec:sage}

After obtaining MPC parameters using the SAGE algorithm as described above, we compute the position of each multipath scattering point, which enables a direct relationship between scattering points and human body parts in space. In this paper, we define a scattering point as the apparent interaction point between the MPC and the human body. It does not necessarily correspond to a physical scatterer, but serves as a virtual point where signal reflection is assumed to occur. Specifically, given the estimated parameter set of the $k$-th path, we denote the MPC parameters as $ \{\tau_k, \theta_k^\mathrm{E}, \theta_k^\mathrm{A}, \alpha_k\} $ representing delay, elevation angle, azimuth angle, and complex amplitude, respectively. Due to the monostatic configuration adopted in our ISAC measurement system, where the transmitter and receiver share the same physical location, we assume a \textit{single bounce propagation model}. Under this assumption, the propagation distance can be directly derived from the time delay. Specifically, for the $k$-th path with estimated delay $\tau_k$, and letting $c_0$ be the speed of light, the propagation distance $d_k$ is calculated as:

\begin{equation}
d_k = \frac{c_0 \cdot \tau_k}{2}
\end{equation}

Assuming the origin is placed at the transmitter/receiver, the scattering point's position $(x_k, y_k, z_k)$ is computed via following transformation, where the elevation angle $\theta_k^\mathrm{E}$ is measured from the horizontal rather than the positive $z$-axis:

\begin{subequations}
\begin{align}
x_k &= d_k \cdot \cos(\theta_k^\mathrm{E}) \cdot \cos(\theta_k^\mathrm{A}) \\
y_k &= d_k  \cdot \cos(\theta_k^\mathrm{E})  \cdot \sin(\theta_k^\mathrm{A}) \\
z_k &= d_k \cdot \sin(\theta_k^\mathrm{E})
\end{align}
\end{subequations}

\begin{figure}
    \centering
    \includegraphics[width=0.35\textwidth]{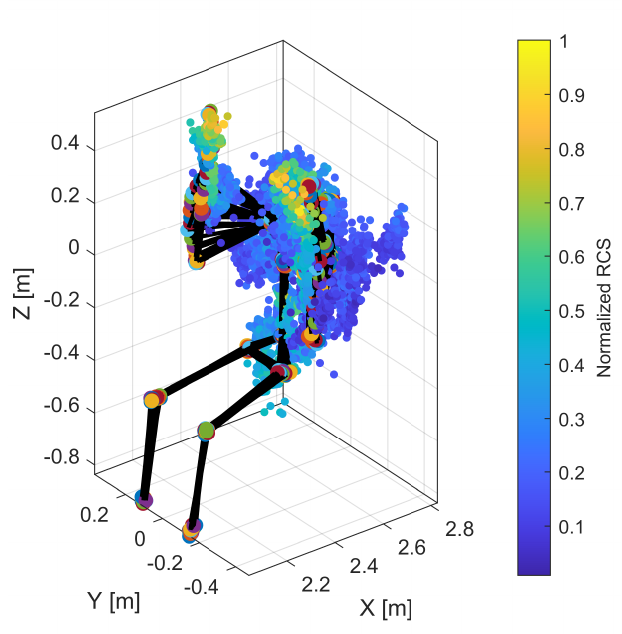} 
    \caption{The results of multipath estimation, aggregated over the entire gesture duration. (Gesture: \textit{Up-Up})}
    \label{fig3}
\end{figure}

The Radar Cross Section (RCS) is an important parameter for characterizing the scattering capability of a target, effectively reflecting the physical properties of the scattering points, which is essential for accurately predicting power in channel modeling. The RCS of the $k$-th path can be approximated using the radar equation\cite{Ref.add1}:
\begin{equation}
\sigma_k = (4\pi)^3 \left( \frac{|\alpha_k| \cdot d_k^2 \cdot f_c}{c_0} \right)^2
\end{equation}
where $|\alpha_k|$ is the amplitude of the $k$-th path estimated from SAGE, $f_c$ is the carrier frequency. Due to the single bounce propagation, the TX and RX distances are approximately equal and can be marked as $d_k$.

The estimated RCS values vary across different body parts and gestures. Under the “\textit{Up-Up}” gesture, the distribution of scattering points estimated using SAGE is shown in Figure~\ref{fig3}. The color of each point represents its normalized RCS. The human stick figure representation is overlaid for reference. It can be observed that stronger scattering points are mainly concentrated around the forearms, suggesting that these regions contribute more significantly to the channel model.

\subsection{Semantic Clustering}
\label{sec:clustering}

To enhance the interpretability of MPCs in the channel, we introduce a semantic clustering method based on human body parts, where the term "semantic" refers to the association of MPCs with physically meaningful interpretations. Unlike conventional clustering methods that rely solely on numerical features, semantic clustering explicitly incorporates domain knowledge of the underlying scenario. Specifically, the estimated scattering points are mapped to different human body regions (e.g., head, torso,  etc.), thereby establishing a direct correspondence between the physical body structure and channel characteristics. This semantic mapping facilitates body-part-level modeling for human gesture channels, which improves the generalization ability, enhances physical consistency, and provides a more structured understanding of MPCs.

To map scattering points to human body parts, we propose a trajectory-based clustering method that combines spatial tracking and distance judgment. First, we track the MPC trajectories using the nearest-neighbor over time, which we normalize so that one unit of time is equal to the sample interval. For each MPC at time $t$, we find the corresponding path at time $t-1$ by minimizing the Euclidean distance between scattering points as follows, where $\|\mathbf{x}\|_2$ is the $L^2$ norm of vector $\mathbf{x}$:
\begin{equation}
j^* = \arg\min_{j \in \{1, \dots, N_c\}} \left\| \mathbf{r}_{n,t} - \mathbf{r}_{j,t-1} \right\|_2
\end{equation}
where $\mathbf{r}_{n,t}$ is the 3D position vector which contains the $(x,y,z)$ coordinates of the $n$-th scattering point at time $t$, $\mathbf{r}_{j,t-1}$ is the position vector of the $j$-th scattering point candidate at time $t{-}1$, $N_c$ is the number of scattering point candidates at time $t{-}1$, and $j^*$ denotes the index of the matched scattering point in the snapshot $t{-}1$.

\begin{figure}
    \centering
    \includegraphics[width=0.35\textwidth]{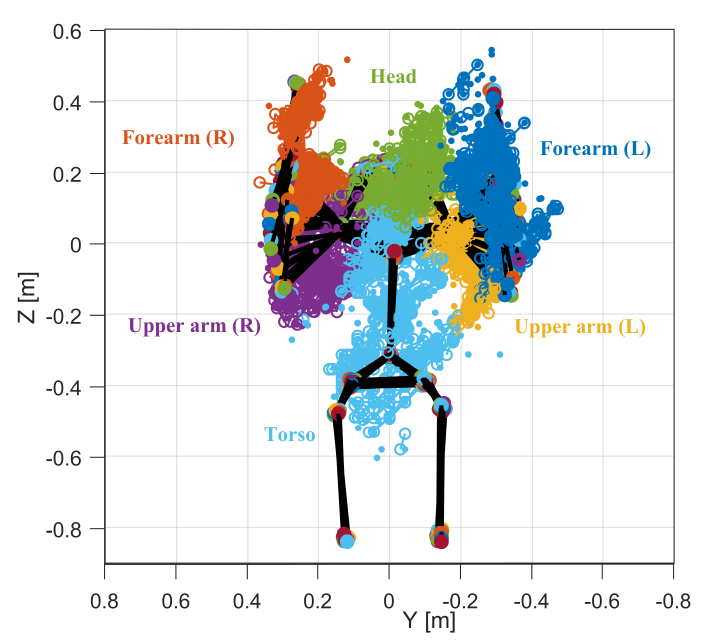} 
    \caption{The results of semantic clustering (Gesture: \textit{Up-Up})}
    \label{fig4}
\end{figure}

Once the MPC trajectories are extracted, we assign each trajectory to a human body part. Instead of labeling scattering points individually, we adopt a trajectory-level decision based on average spatial proximity to predefined body parts. Specially, let $T_k = \{\mathbf{r}_1, \mathbf{r}_2, \dots, \mathbf{r}_N\}$ denote the $k$-th MPC trajectory consisting of $N$ scattering points. As shown in Figure~\ref{fig2}, we define human body parts $\{ \mathcal{S}_1, \mathcal{S}_2, \dots, \mathcal{S}_6 \}$, each represented by one or more line segments connecting two keypoints, as listed in Table I. The body part label $\hat{L}_k$ of trajectory $T_k$ is:
\begin{equation}
\hat{L}_k = \arg\min_{j \in \{1,\dots,6\}} \left( \frac{1}{N} \sum_{n=1}^{N} \text{Dist}(\mathbf{r}_n, \mathcal{S}_j) \right)
\end{equation}
where $\text{Dist}(\mathbf{r}_n, \mathcal{S}_j)$ denotes the shortest distance from the $n$-th scattering point with position $\mathbf{r}_n$ to the center of the $j$-th body part $\mathcal{S}_j$.
In addition, outlier scattering points that exceed a predefined threshold distance (set as 25~cm in this paper) from all body parts are filtered out. Figure~\ref{fig4} presents an example of the semantic clustering results under the “\textit{Up-Up}” gesture. It can be observed that the clustering results exhibit spatial separation among different regions, such as the head, torso, upper arms, and forearms.

\begin{table}[ht]
\centering
\caption{Mapping between Human Body Parts and keypoints}
\label{tab:body_parts}
\begin{tabular}{cc}
\toprule
\textbf{Human Body Parts} & \textbf{Connected keypoints} \\
\midrule
Forearm (L)   & Left Elbow – Left Wrist \\
Forearm (R)   & Right Elbow – Right Wrist \\
Upper Arm (L) & Left Shoulder – Left Elbow \\
Upper Arm (R) & Right Shoulder – Right Elbow \\
Head          & Nose – Left Eye – Right Eye \\
              &  – Left Ear – Right Ear \\
Torso         & Left Shoulder – Right Shoulder – Left Hip \\
              &  – Right Hip – Belly – Chest\\
\bottomrule
\end{tabular}
\end{table}

\section{DL-based Channel Modeling Approach}
\label{sec:classification}

Deep learning approach is an important subfield of machine learning, centered on building multi-layer neural networks to automatically extract high-level features from data \cite{Ref35}. In this section, we present the details of the DL-based channel modeling approach, which aims to simulate channel characteristics that correspond to human gesture variations.

\subsection{Overview}

The scattering points under different human gestures exhibit different spatial distribution characteristics, which should be exploited by deep learning, as detailed below:

\begin{itemize}
    \item \textbf{Randomness:} Due to random subtle movements and noise, the number of scattering points observed at different snapshots do not follow a deterministic mapping, but rather a stochastic process statistically correlated with the human gesture.

    \item \textbf{Sparsity:} Each snapshot contains only a few dominant scattering points, and some body parts may lack sufficient multipath information.

    \item \textbf{Spatial mapping:} The scattering points are not distributed uniformly in 3D space, but are instead highly concentrated around key body parts such as the arms, head, and torso.
\end{itemize}

\begin{figure}
    \centering
    \includegraphics[width=0.45\textwidth]{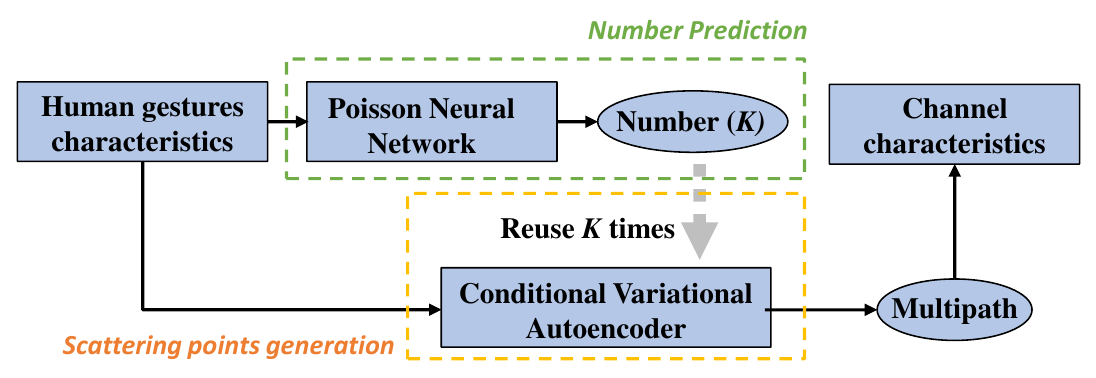} 
    \caption{An overview of DL-based channel modeling approach}
    \label{fig5}
\end{figure}

Given the above characteristics, the scattering points exhibit significant imbalance across different human body parts (e.g., there may be no scattering points in some snapshots), and the number of scattering points associated with similar gestures may vary considerably. The conventional end-to-end deep learning models struggle to handle such randomness and sparsity. To address this, we propose a two-stage generative framework based on deep learning, as illustrated in Figure~\ref{fig5}. The framework decouples the modeling into two sub-tasks: (i) predicting the number of paths and (ii) generating scattering points, which improves both flexibility and interpretability.

Firstly, we employ a Poisson neural network to predict the number of MPCs $K$ within each human body part, based on the human gesture characteristics of the current snapshot. Since the output of a Poisson neural network is a count variable sampled from a Poisson distribution, this approach enables modeling of uncertainty. For each region, a C-VAE is reused $K$ times to generate scattering points, including position and RCS parameters. Finally, based on the generated scattering points, the channel impulse response is reconstructed, enabling the simulation of key channel characteristics in the human gestures, including path gain, delay, and micro-Doppler shift.

\subsection{MPC Number Prediction}
Poisson neural networks are a class of models designed to predict count-based outputs that follow a Poisson distribution \cite{Ref36}. Unlike traditional regression networks that produce continuous values, Poisson neural networks output the rate parameter $\lambda$ of a Poisson distribution, from which discrete, non-negative counts can be sampled. This makes them particularly suitable for modeling event occurrences or component counts in sparse and stochastic systems.

\begin{figure}
    \centering
    \includegraphics[width=0.4\textwidth]{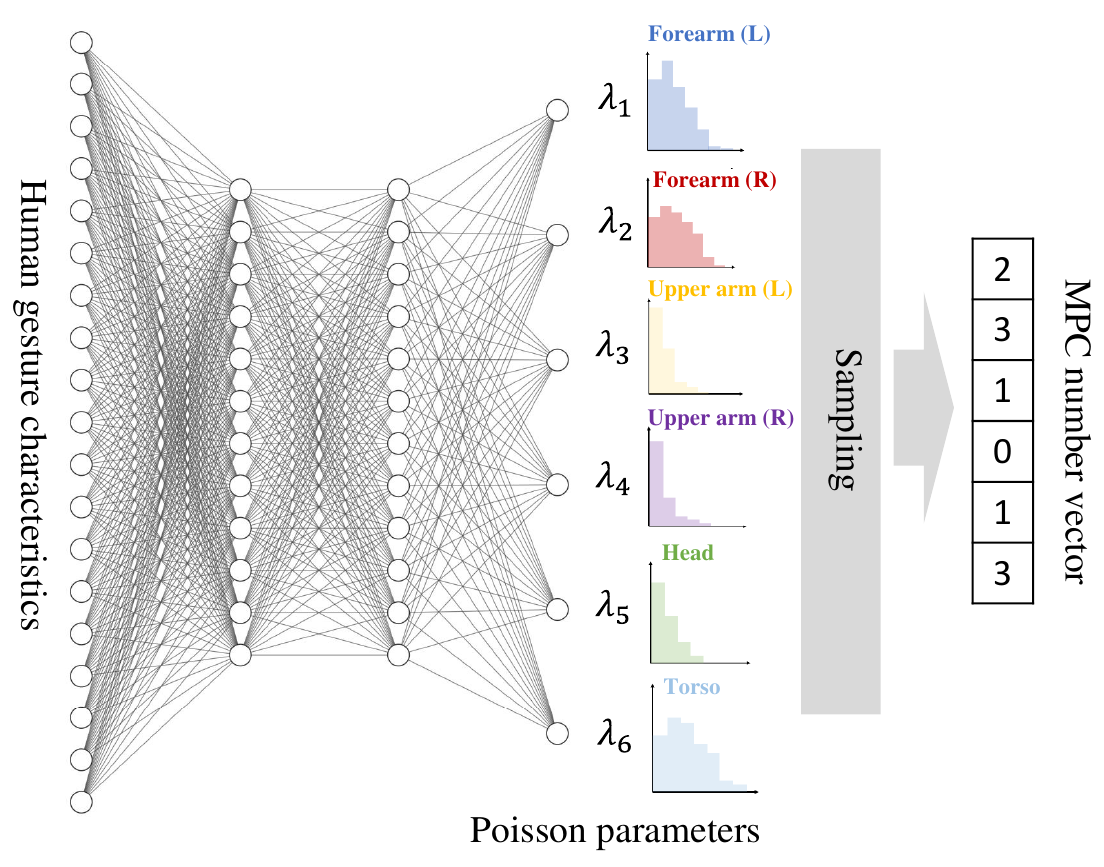} 
    \caption{The MPC number prediction based on Poisson neural networks}
    \label{fig6}
\end{figure}

To predict the number of MPCs in different human body parts, we design a Poisson neural network using a multi-layered architecture, as shown in Figure~\ref{fig6}. The input is a human gesture characteristics matrix with 19 keypoints in 3D space (dimension $19 \times 3$), and the output is a vector of Poisson rate parameters $\boldsymbol{\lambda} = [\lambda_1, \lambda_2, \dots, \lambda_6]$, corresponding to the six human body parts that are developed from the keypoints. The final output layer uses an exponential activation function $\exp(\cdot)$ to ensure all $\lambda_i$ values are positive, thus satisfying the requirement of Poisson distributions. The model is trained by minimizing the Poisson Negative Log-Likelihood (NLL) loss between the predicted parameters $\lambda_{ij}$ and ground truth number $k_{ij}$, which is shown as follows:
\begin{equation}
\mathcal{L}_{\text{Poisson}} = \frac{1}{M} \sum_{m=1}^{M} \sum_{j=1}^{6} \left( \lambda_{mj} - k_{mj} \log (\lambda_{mj} + \epsilon) \right)
\end{equation}
where $\lambda_{ij}$ denotes the predicted Poisson rate for the $j$-th body part in the $m$-th sample, $M$ is the batch size and $\epsilon$ is a small constant for numerical stability. The loss is minimized using the Adaptive Moment Estimation (Adam) algorithm, a commonly used method for adjusting model parameters\cite{Ref.add5}.

During inference, the predicted Poisson rate vector $\boldsymbol{\lambda} = [\lambda_1, \lambda_2, \dots, \lambda_6]$ is used to sample the number of MPCs in each human body part as follows:
\begin{equation}
K_j \sim \mathrm{Poisson}(\lambda_j), \quad j = 1, 2, \dots, 6.
\end{equation}
This statistically distributed modeling neural network introduces uncertainty and randomness into the subsequent generative process, making the channel simulation more realistic.

\begin{figure*}[!ht]
    \centering
    \includegraphics[width=0.8\textwidth]{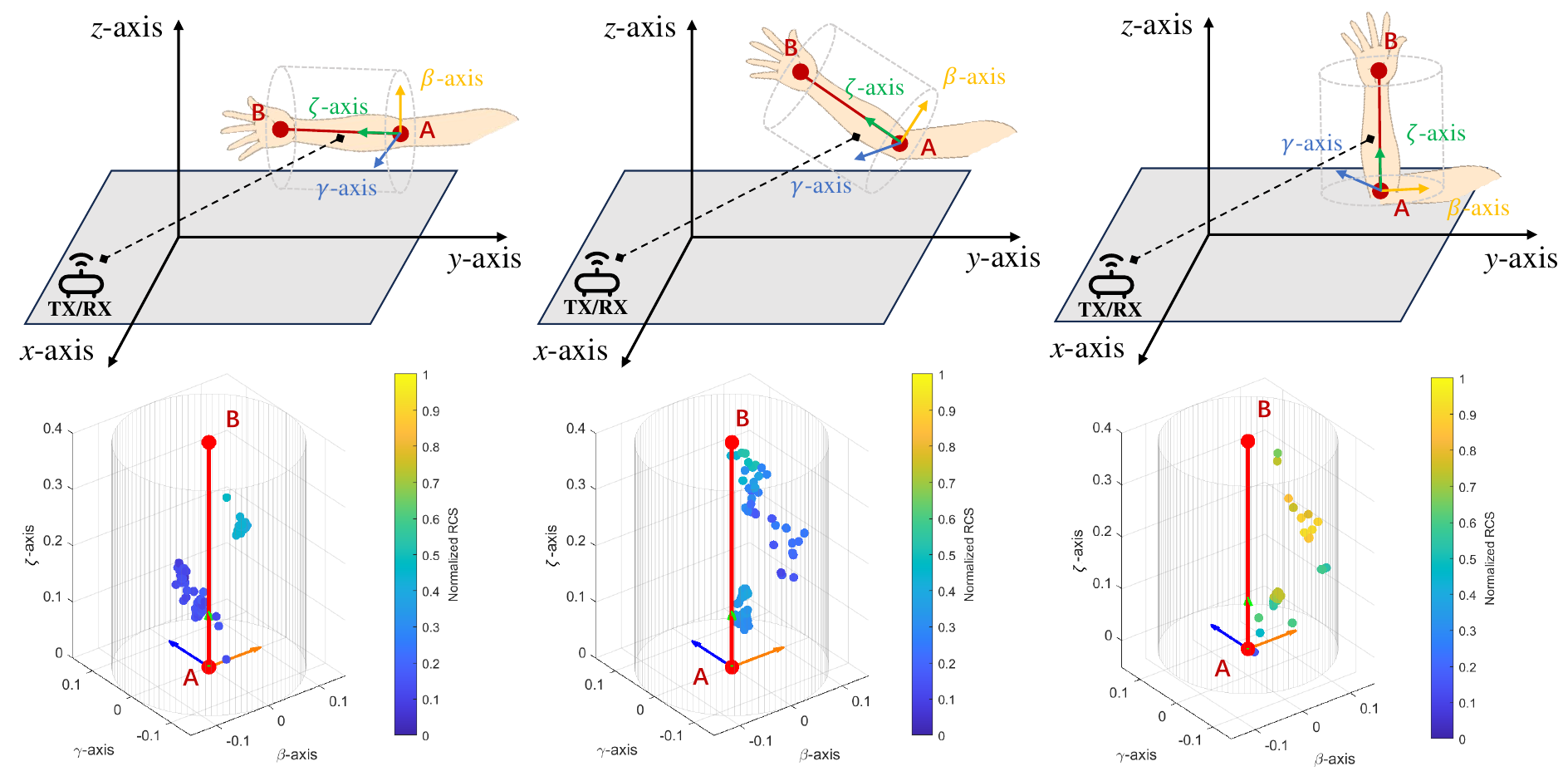} 
    \caption{Local coordinate systems and multipath distributions under time-varying forearm}
    \label{fig7}
\end{figure*}

\subsection{Scattering Points Generation}

In the scattering points generation sub-task, we first introduce a local coordinate normalization mechanism. Compared to directly using global coordinates, the local coordinate system is constructed based on related two keypoints, which better captures the scattering characteristics within local regions while reducing interference caused by global gesture variations. Although this introduces additional computational loads for coordinate transformation, it significantly improves the stability and generalization of neural networks.

Specifically, as shown in Figure~\ref{fig7}, we define a local coordinate system $(\zeta,\beta,\gamma)$ for each human body part based on the associated keypoints A, B with position vectors $\mathbf{r}_\mathrm{A}$, $\mathbf{r}_\mathrm{B}$, respectively, and the collocated position of the TX/RX ($\mathbf{r}_\mathrm{TX}$). With the forearm moving, the scattering points within the region exhibit learnable and predictable patterns. At the first snapshot, the origin of the local coordinate system is set to point A. The $\zeta$-axis is defined through the unit vector $\hat{\boldsymbol{\zeta}}$ from A to B, while the unit vector of $\gamma$-axis ($\hat{\boldsymbol{\gamma}}$) is defined as the normalized transverse component of the vector from A to TX with respect to $\hat{\zeta}$. The unit vector of $\beta$-axis is finally computed as the cross product between $ \hat{\boldsymbol{\zeta}}$ and $\hat{\boldsymbol{\gamma}}$, i.e. $\hat{\boldsymbol{\beta}}= \hat{\boldsymbol{\gamma}} \times \hat{\boldsymbol{\zeta}}$. As a result, we obtain the right-handed orthonormal local coordinate system depicted in Figure~\ref{fig7}, and represented by the rotation matrix:
\begin{equation}
\underline{\mathbf{R}} = [\hat{\boldsymbol{\zeta}}, \hat{\boldsymbol{\beta}}, \hat{\boldsymbol{\gamma}}]
\end{equation}

In subsequent snapshots, the positions of keypoints A and B change over time, then the vector triplet $\left(\hat{\boldsymbol{\zeta}}_t, \hat{\boldsymbol{\beta}}_t, \hat{\boldsymbol{\gamma}}_t\right)$ and the rotation matrix $\underline{\mathbf{R}}_t$ are recalculated at each time $t$ to ensure that the local coordinate system remains aligned with the orientation of the associated body part. 

Additionally, at each time $t$ we transfer the global coordinates $\mathbf{r}_{n,t}$ of the $n$-th scattering point into the local frame of the corresponding body part, obtaining the local position vector $\mathbf{r}^\mathrm{local}_{n,t}$, with origin in the keypoint A:
\begin{equation}
\mathbf{r}^\mathrm{local}_{n,t} = \underline{\mathbf{R}}_t \cdot \left( \mathbf{r}_{n,t} - \mathbf{r}_{\mathrm{A},t} \right)
\end{equation}
Based on the local coordinates, we further introduce a C-VAE to generate the scattering points within each human body part. The C-VAE is a classical deep generative model that extends the VAE by incorporating conditional modeling \cite{Ref37}. By introducing additional conditioning information $c$ into both the encoder and decoder, C-VAE enables the generation process to be guided by prior knowledge. 

The overall architecture of the C-VAE consists of three components: an encoder $q_{\varphi}(z|x, c)$, a decoder $p_{\Omega}(x|z, c)$, and a conditional prior $p(z|c)$. Here, $\varphi$ and $\Omega$ denote the learnable parameters of the encoder and decoder networks, respectively. The encoder maps the input scattering points feature $x$ and conditional label $c$ to a latent variable $z$ modeled as a Gaussian distribution, with parameters $\mu$ and $\sigma^2$ for the mean and variance. To enable backpropagation through the sampling operation, the reparameterization trick is used during training:
\begin{equation}
z = \mu_{\varphi} + \sigma_{\varphi} \cdot \epsilon, \quad \epsilon \sim \mathcal{N}(0, I)
\end{equation}
The decoder then reconstructs the input feature $x$ from the sampled latent variable $z$ and the same condition $c$. This structure enables the model to learn the conditional distribution of scattering points given the physical and temporal features.

\begin{figure}
    \centering
    \includegraphics[width=0.5\textwidth]{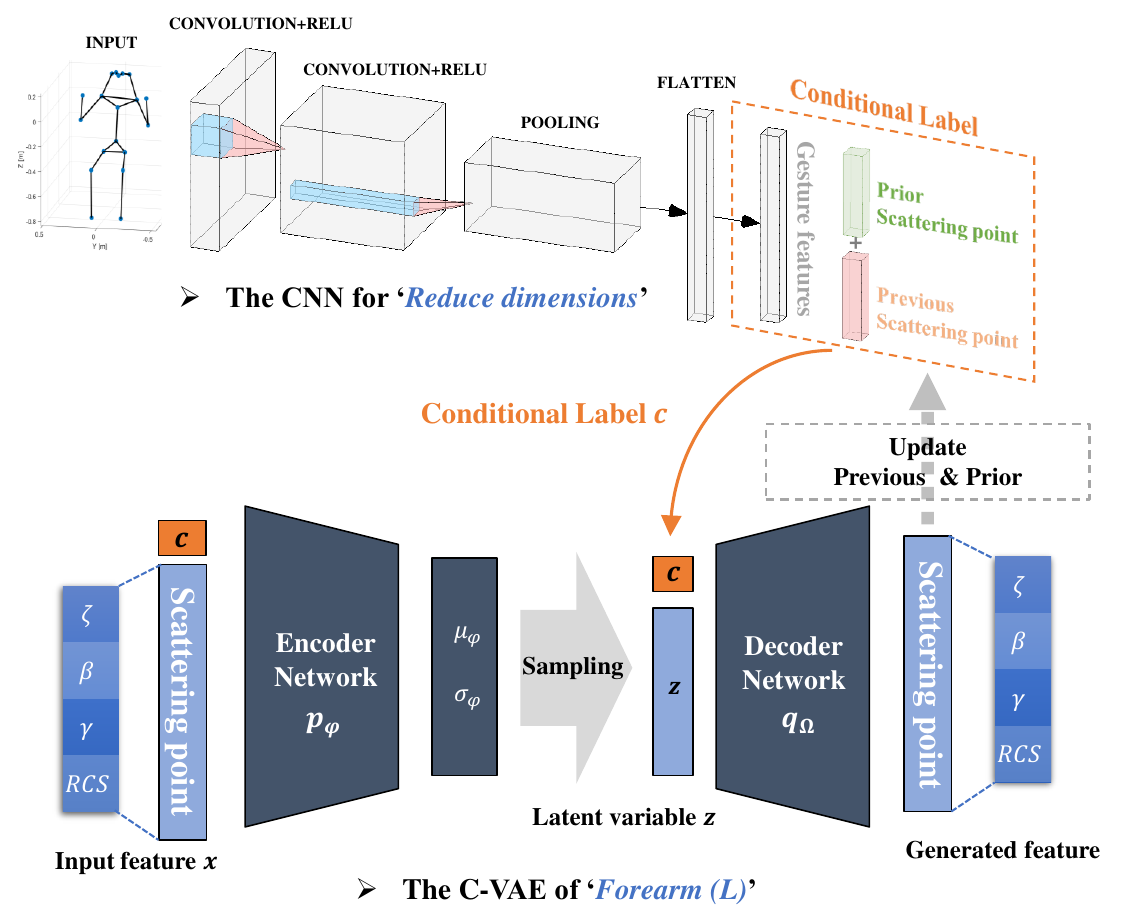} 
    \caption{The scattering points generation based on conditional variational autoencoders}
    \label{fig8}
\end{figure}

In the proposed method, an individual C-VAE is trained for each human body part to accurately model the spatial distribution of scattering points. As shown in Figure~\ref{fig8}, for each snapshot, a lightweight Convolutional Neural Network (CNN) is first employed to extract a low-dimensional gesture feature vector $\mathcal{C}_{\text{ges}}$ (dimension $8 \times 1$) from the 3D coordinates of human body keypoints (dimension $19 \times 3$).
To further enhance the temporal and spatial consistency in scattering point generation, we introduce two auxiliary conditional labels for each target scattering point, based on the associated time-variant MPC. They are dynamically updated as each new scattering point is generated, enabling the model to maintain spatial and temporal consistency across paths and snapshots, which are introduced as follows:

\begin{itemize}
\item $\mathcal{C}_{\text{pri}}$: the feature vector of the scattering point ranked just before the current one -- called the \textit{prior} scattering point -- based on the delay of the associated MPC within the same time snapshot (i.e., the spatially adjacent path with shorter delay than the current one).
\item $\mathcal{C}_{\text{pre}}$: the feature vector of the temporally matched scattering point from the \textit{previous} time snapshot, representing the same path across time. The match is computed based on the results of Section~\ref{sec:clustering}.
\end{itemize}

These three vectors are concatenated to form the complete conditional label $\mathcal{C} = \{\mathcal{C}_{\text{ges}}, \mathcal{C}_{\text{pri}},\mathcal{C}_{\text{pre}}\}$, which represent gesture-level features, prior intra-frame path information, and inter-frame temporal information, respectively. The conditional label is input to the C-VAE along with the scattering point feature, which includes the local coordinates $(\zeta, \beta, \gamma)$ and RCS. At the beginning of the generation process, both $\mathcal{C}_{\text{pri}}$ and $\mathcal{C}_{\text{pre}}$ are initialized as zero vectors, which have the same dimensionality as the input scattering point feature. During inference, the C-VAE is reused $K$ times for each human body part to generate $K$ scattering points, where $K$ is the number of MPCs predicted by the Poisson neural network. After each scattering point is generated, $\mathcal{C}_{\text{pre}}$ and $\mathcal{C}_{\text{pri}}$ are updated, guiding the generation of the next scattering point. By learning temporal and spatial dependencies, the framework ensures consistent and continuous scattering point generation. The generated scattering points from all body parts are then aggregated to reconstruct the human gesture channels.

\subsection{Channel Modeling}

To reconstruct the human gesture channel characteristics based on the generated scattering points, we simulate both the CIR and the micro-Doppler signature, which are crucial for human gesture recognition. Each generated scattering point is first mapped to global coordinates using the TX/RX location and its local coordinates $(\zeta, \beta, \gamma)$, along with its corresponding two keypoints. Based on global coordinates, the total propagation distance is computed from the spatial positions of the TX/RX and the $n$-th scattering point. The corresponding propagation delay $\tau_{n,t}$ at time $t$ is then calculated as:
\begin{equation}
\tau_{n,t} = \frac{2 \cdot d_{n,t}}{c_0}
\end{equation}
where $d_{n,t}=\|\mathbf{r}_\mathrm{TX}-\mathbf{r}_{n,t}\|$ is the Euclidean distance between the transceiver and the $n$-th scattering point in the $t$-th snaphot.
The path gain $|\alpha_{n,t}|^2$ is calculated via the radar equation:
\begin{equation}
|\alpha_{n,t}|^2 = \frac{c_0^2 \cdot \sigma_{n,t}}{(4\pi)^3 \cdot f_c^2 \cdot d^{4}_{n,t}}
\end{equation}
where $\sigma_{n,t}$ is the RCS of the $n$-th scattering point. 

To simulate the micro-Doppler shift, we assume that each human body part undergoes rigid body motion, defined by the keypoints A and B, and the $n$-th scattering point is rigidly attached to this region. According to the Mozzi-Chasles theorem, the most general (instantaneous) displacement for a rigid body is a screw displacement, and the instantaneous velocities of each pair of points (say A and $n$) satisfy the rigid body velocity law \cite{Rigid_body}:
\begin{equation}
\mathbf{v}_{n,t} = \mathbf{v}_{\mathrm{A},t} + \boldsymbol{\omega}_t \times (\mathbf{r}_{n,t} - \mathbf{r}_{\mathrm{A},t})
\label{rigid_body_law1}
\end{equation}
with $\mathbf{r}_{\mathrm{A},t},\,\mathbf{v}_{\mathrm{A},t}$ position and velocity of the keypoint A at time $t$, respectively, while $\mathbf{r}_{n,t},\,\mathbf{v}_{n,t}$ are the instantaneous position and velocity of the $n$-th scattering point. The angular velocity vector of the rigid body is $\boldsymbol{\omega}_t=\omega_t\,\hat{\mathbf{k}}_t$
with $\omega_t$ instantaneous angular frequency, while the unit vector $\hat{\mathbf{k}}_t$ gives the direction of the screw axis at time $t$.

Similarly, for points B and $n$ the following relation holds:
\begin{equation}
\mathbf{v}_{n,t} = \mathbf{v}_{\mathrm{B},t} + \boldsymbol{\omega}_t \times (\mathbf{r}_{n,t} - \mathbf{r}_{\mathrm{B},t})
\label{rigid_body_law2}
\end{equation}

Then, the angular velocity vector $\boldsymbol{\omega}_t$ can be determined based on the known instantaneous velocities and positions of keypoints A and B,
by equating \eqref{rigid_body_law1} and \eqref{rigid_body_law2} and solving the equation with respect to $\boldsymbol{\omega}_t$ \cite{Ref.add2}:
\begin{equation}
\boldsymbol{\omega}_t = \frac{(\mathbf{r}_{\mathrm{B},t} - \mathbf{r}_{\mathrm{A},t}) \times (\mathbf{v}_{\mathrm{B},t} - \mathbf{v}_{\mathrm{A},t})}{\|\mathbf{r}_{\mathrm{B},t} - \mathbf{r}_{\mathrm{A},t}\|^2}
\label{rigid_body_law3}
\end{equation}

Subsequently, the instantaneous velocity $\mathbf{v}_{n,t}$ of the $n$-th scattering point is obtained using \eqref{rigid_body_law1} or \eqref{rigid_body_law2}. 

It is worth noting that \eqref{rigid_body_law3} is obtained under the assumption $\left[\boldsymbol{\omega}_t \cdot \left( \mathbf{r}_{\mathrm{B},t} - \mathbf{r}_{\mathrm{A},t} \right)\right]=0$, which means that the screw axis is always perpendicular to the $\zeta$ axis, and then any rolling displacement around $\zeta$ is neglected. This assumption is generally valid for most human gestures, but even if some rolling displacement were present, it would have a negligible influence on the micro-Doppler effect.

The Doppler frequency shift induced by the motion of each scattering point then follows as:
\begin{equation}
\Delta f_{n,t} = \frac{2 f_c}{c_0}\left( \mathbf{v}_{n,t} \cdot \frac{\mathbf{r}_\mathrm{TX}-\mathbf{r}_{n,t}}{\|\mathbf{r}_\mathrm{TX}-\mathbf{r}_{n,t}\|}\right) 
\end{equation}
where $\mathbf{r}_\mathrm{TX}-\mathbf{r}_{n,t}$ is the vector from the $n$-th scattering point to the transceiver at $t$, and $\mathbf{v}_{n,t}$ is the velocity of the scattering point at $t$. Based on this formulation, the channel impulse response is then modeled as the summation of all generated MPCs, as follows:
\begin{equation}
h_t(\tau) = \sum_{n=1}^{N_t} \alpha_{n,t} \, e^{j\big[\psi_n + 2\pi (f_c \tau_{n,t}+\Delta f_{n,t}t)\big]} \, \delta(\tau - \tau_{n,t})
\end{equation}
where $\psi_n \sim \mathcal{U}[-\pi, \pi]$ is a uniformly distributed random phase and $N_t$ is the total number of generated scattering points at time $t$ (i.e., $N_t = \sum_{j=1}^{6} K_{j,t}$), and $\delta(\cdot)$ represents the Dirac delta function. Next, to generate the micro-Doppler frequency power spectrum, we compute the short-time Fourier transform over period $T$:
\begin{equation}
H_t(\Delta f)= \int^{t+T/2}_{t-T/2} h_{\tilde{t}}(\tau)e^{-j2\pi \Delta f \tilde{t}} d\tilde{t}
\end{equation}
By setting $T=0$, we obtain the {\em instantaneous} spectrum as:
\begin{eqnarray}
S_t(\Delta f) &=& H_t(\Delta f) \cdot H_t^*(\Delta f) \nonumber \\&=&\sum_{n=1}^{N_t}|\alpha_{n,t}|^2 \cdot \delta(\Delta f - \Delta f_{n,t}),
\end{eqnarray}
where * denotes the complex conjugate.
This formulation reflects the dynamic frequency contributions of moving scattering points.

\section{Neural Network Training and Performance}

This section describes the training and evaluation of the proposed neural networks, including the configuration of hyper-parameters, prediction performance, and scattering point generation ability. Specifically, subsection A summarizes the training settings for both the Poisson neural network and the C-VAE. Subsections B and C then evaluate the prediction accuracy of the Poisson neural network and the generation performance of the C-VAE, respectively.

\begin{table}[ht]
\centering
\caption{Dataset Partition and Hyper-parameter Configuration}
\label{tab:hyperparameters}
\renewcommand{\arraystretch}{1.2}
\setlength{\tabcolsep}{4pt}  
\begin{tabular}{>{\centering\arraybackslash}p{1.3cm} 
                >{\centering\arraybackslash}p{2.5cm} 
                >{\centering\arraybackslash}p{4.0cm}}
\toprule
\textbf{Category} & \textbf{Item} & \textbf{Configuration} \\
\midrule
\multirow{2}{*}{Dataset} 
& Training Set & 3 subjects \\
& Testing Set & 1 subject \\
\midrule
\multirow{6}{*}{Poisson Net}
& Input Shape & (19, 3) \\
& Output Size & (6, 1) \\
& Dense Layers & 512, 128, 6 \\
& Activation & ReLU + Exponential \\
& Epochs & 250 \\
& Batch &  64 \\
& Learning Rate & 0.001 (Adam optimizer) \\
\midrule
\multirow{9}{*}{C-VAE}
& Input Shape & (19, 3)  \\
& Output Size & (4, 1) \\
& Condition Vector & Gesture feature (8) + \newline previous path (4) + prior path (4) \\
& Latent Dim & 10 \\
& Convolutional Filters & 16 \\
& Convolutional Filter Size & (5, 1) \\
& Encoder Layers & 256, 128, 64 \\
& Decoder Layers & 64, 128, 256 \\
& Epochs & 250 \\
& Batch & 64 \\
& Learning Rate & 0.001 (Adam optimizer) \\
\bottomrule
\end{tabular}
\end{table}

\subsection{Neural Network Hyper-parameters}

The hyper-parameter configurations are summarized in Table II, which can be categorized into dataset settings, model structure (e.g., layer size, activation), and training configuration (e.g., learning rate, batch size). For each dataset, three subjects are used for training and the remaining one is used for testing.
The Poisson neural network is a multi-layer network with Rectified Linear Unit (ReLU) activation functions between layers and an exponential output activation function, while the C-VAE incorporates convolutional layers and dense layers for encoding and decoding, conditioned on gesture and path-related features.

During the training process, both models are optimized using the Adam algorithm. The learning rate, which controls how much the model updates during each step, is set to 0.001. Training is performed in small groups of 64 samples at a time (batch size = 64), and the entire dataset is processed 250 times (known as epochs) to ensure convergence.

\subsection{Prediction performance}

To evaluate the prediction performance of the Poisson neural network, we compare the predicted and true distributions of MPC number at different snapshots in the testing set. For measured data, considering the snapshot interval is 2.6~ms, we assume the human gesture remains unchanged over 10 adjacent snapshots and count the number of MPCs within this window to approximate the true distribution. For the predicted data, we repeat the sampling process 100~times using the same gesture feature input to get the distribution obtained by the Poisson neural networks.

\begin{figure}
\centering
\subfigure[]{\includegraphics[width=1.7in]{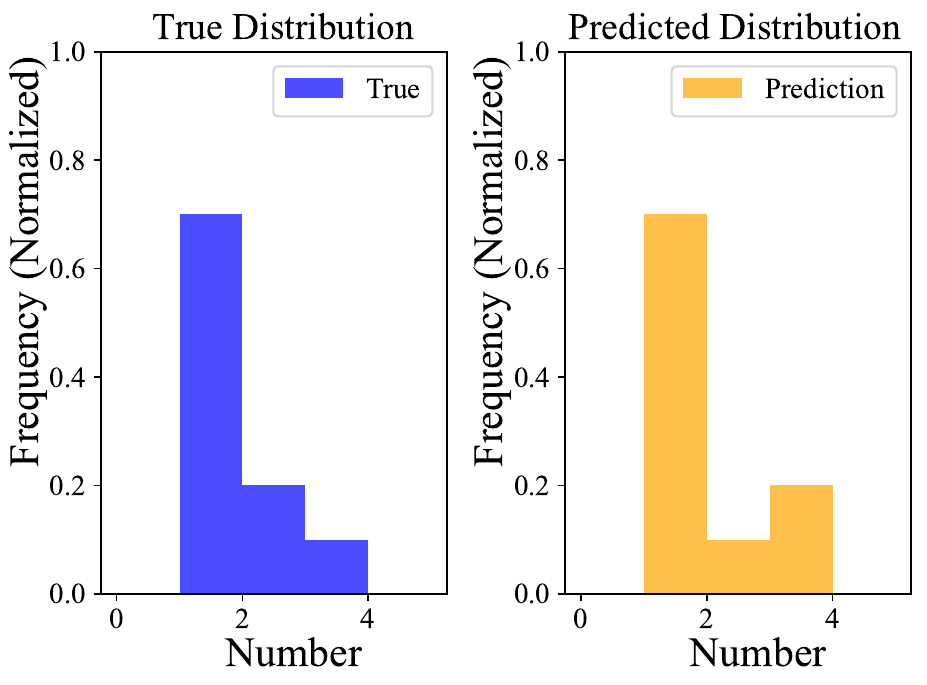}}
\subfigure[]{\includegraphics[width=1.7in]{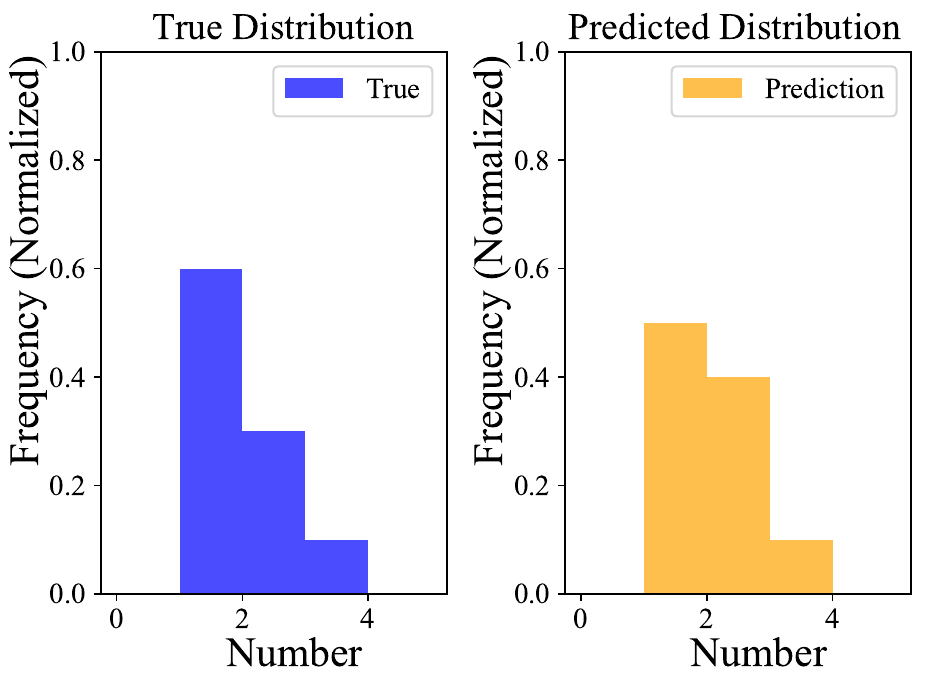}}
\subfigure[]{\includegraphics[width=1.7in]{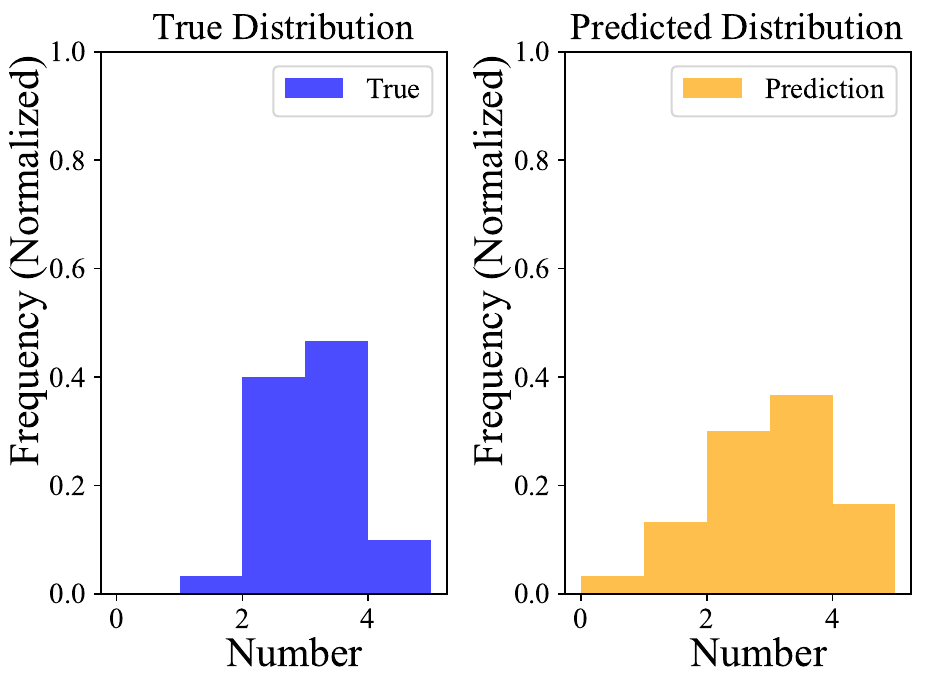}}
\subfigure[]{\includegraphics[width=1.7in]{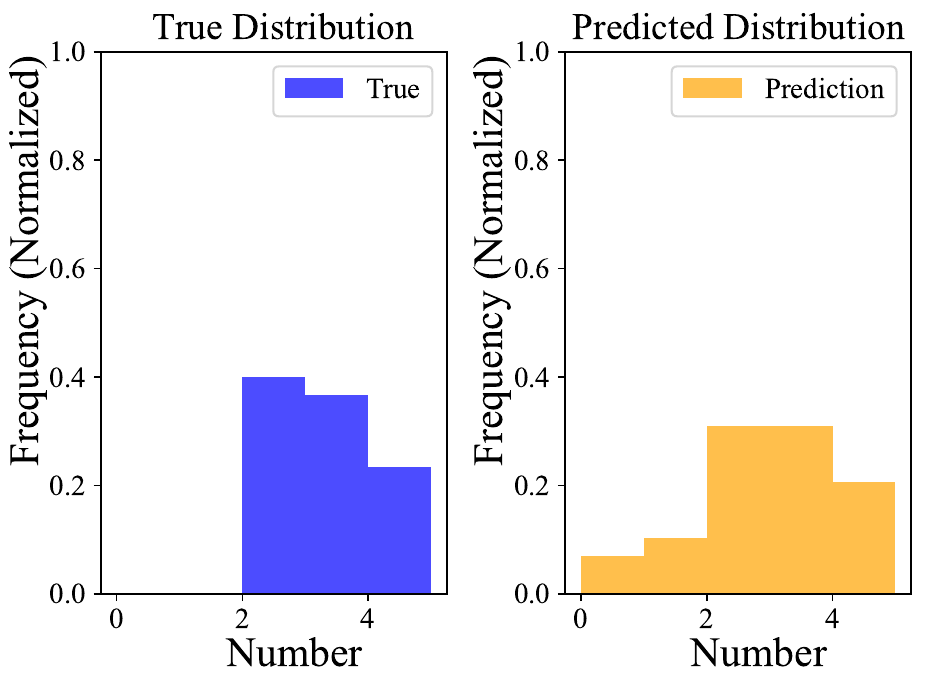}}
\caption{The results of MPC number prediction. (a) First second,  forearm (L). (b) Third second, forearm (L). (c) First second, torso. (b) Third second, torso.}
\label{fig9}
\end{figure}

Figure~\ref{fig9} presents representative results for different human body parts (Forearm and Torso) and snapshots (1st and 3rd seconds). It can be observed that the MPC number distribution varies over time as the gesture changes. Meanwhile, the Poisson neural network successfully captures these changes, and the predicted distributions closely follow the trends of the true number. This demonstrates that the proposed model not only preserves the randomness but also provides time-varying distributional fitting across different body parts.

\subsection{Generation performance}

\begin{figure}
\centering
\subfigure[]{\includegraphics[width=3in]{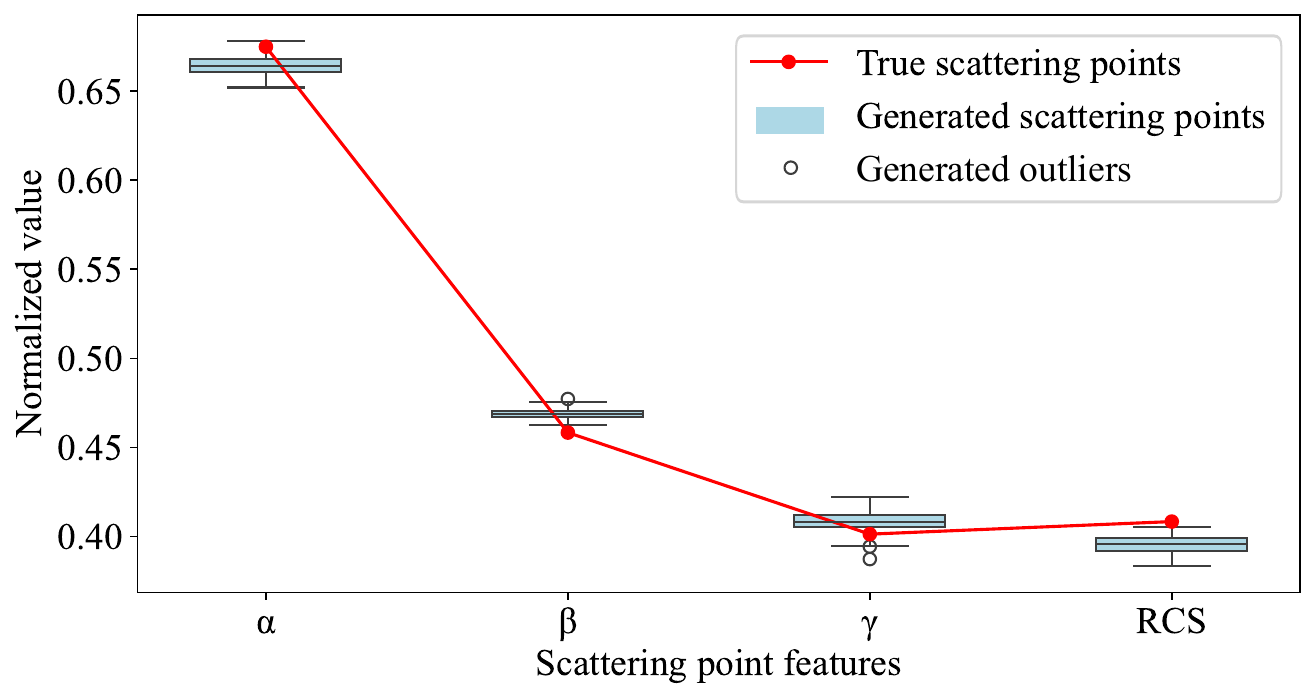}}
\subfigure[]{\includegraphics[width=3in]{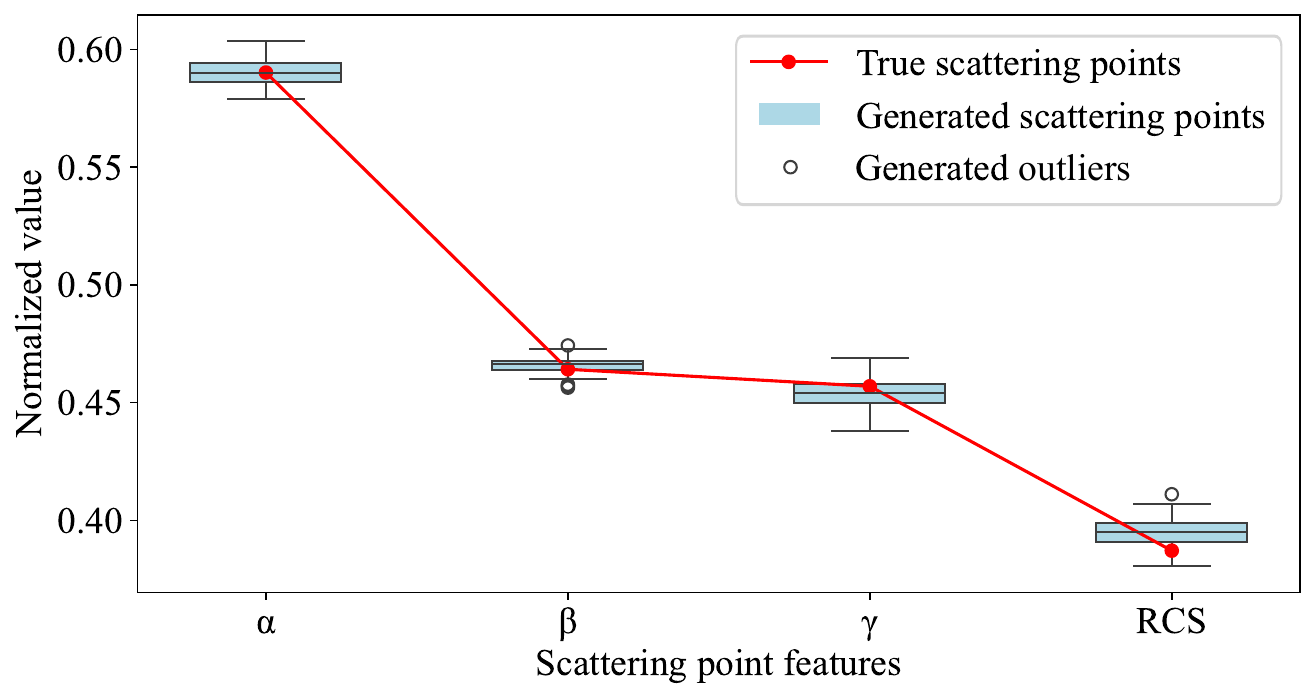}}
\caption{Results of scattering points generation. (a) First second. (b) Third second. Red lines indicate the true scattering point features, while gray boxplots represent the distribution of generated scattering point features. Each boxplot shows the 25th, 50th (median), and 75th percentiles of the generated data, with whiskers extending to 1.5 times the interquartile range.}
\label{fig10}
\end{figure}

To evaluate the ability of the C-VAE model to generate scattering points, we select a specific human gesture from the testing set and compare the generated scattering point features  with those of the real scattering points. Figure~\ref{fig10} presents the results between generated and real scattering points in the forearm, at both the first and third second. The comparison is conducted across four features: local spatial coordinates $(\zeta, \beta, \gamma)$ and RCS. Due to the randomness of scattering points, a one-to-one correspondence at the path-level cannot be guaranteed. Instead, we focus on whether the generated scattering points statistically match the real ones. Figure~\ref{fig10} indicates that the scattering points generated by the C-VAE can effectively match the distributional trend of the real data, capturing the statistical properties under different snapshots. This validates the effectiveness of the proposed conditional generation methods in enhancing the diversity and consistency of scattering points modeling.

\section{Channel Simulation Results}

This section presents the simulation results for the proposed human-gesture-driven channel modeling framework, including cross-validation on unseen gestures and new subjects, as well as the reconstructed CIR and micro-Doppler signatures derived from the generated scattering points. These results are used to evaluate the model's performance in terms of generalization capability, physical consistency, and adaptability to different scenarios.

\begin{figure}
    \centering
    \subfigure[]{\includegraphics[width=1.45in]{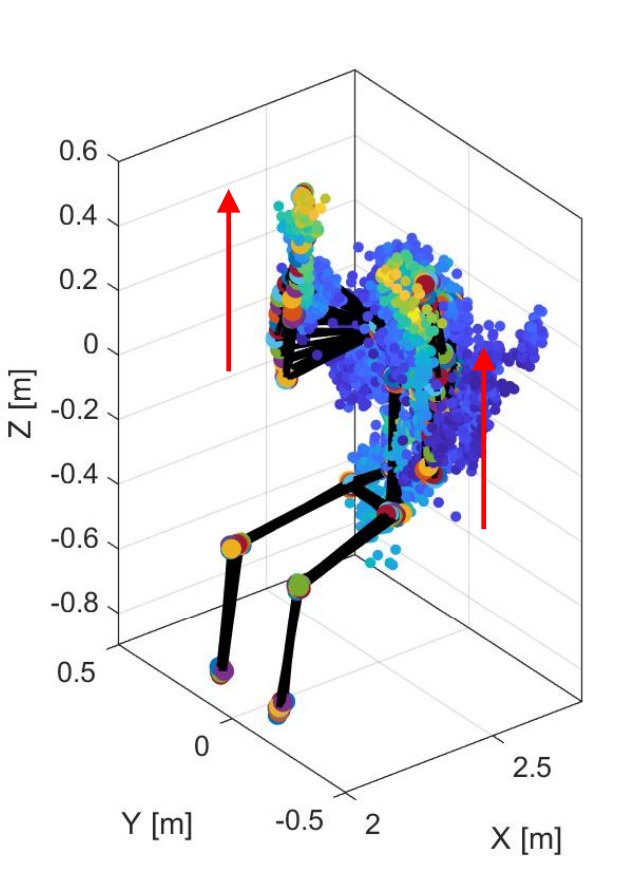}}
    \subfigure[]{\includegraphics[width=1.9in]{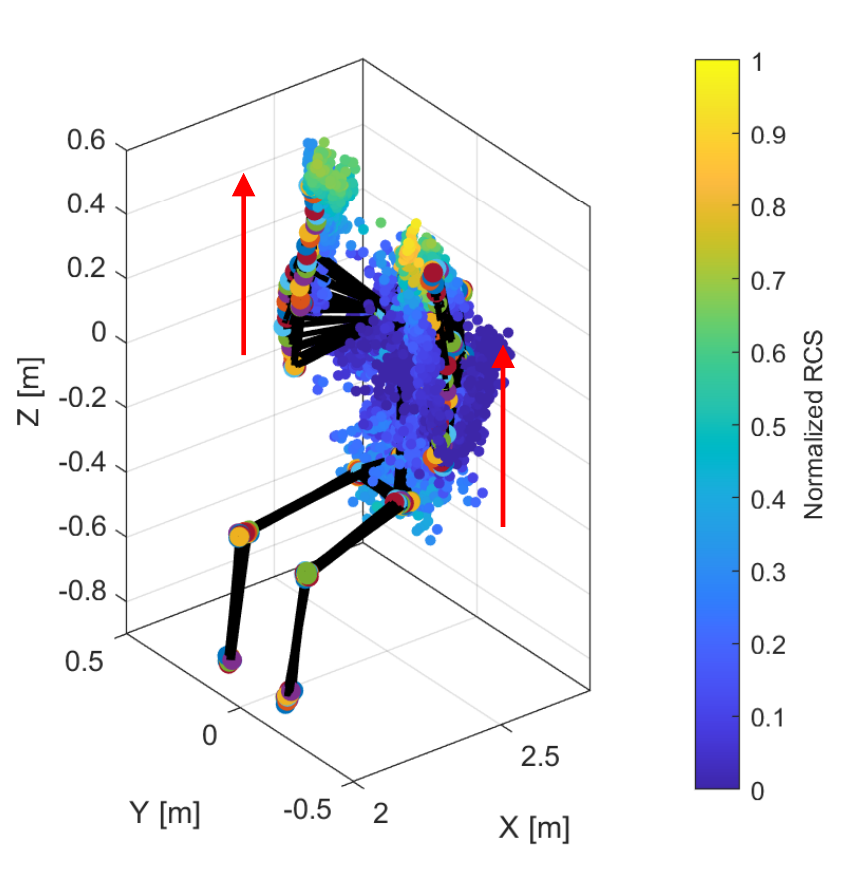}}
    \caption{Comparison between measured and simulated scattering points on a new subject, aggregated over the entire gesture duration. (a) Measured. (b) Simulated.}
    \label{fig11}
\end{figure}

\begin{figure}
    \centering
    \subfigure[]{\includegraphics[width=1.45in]{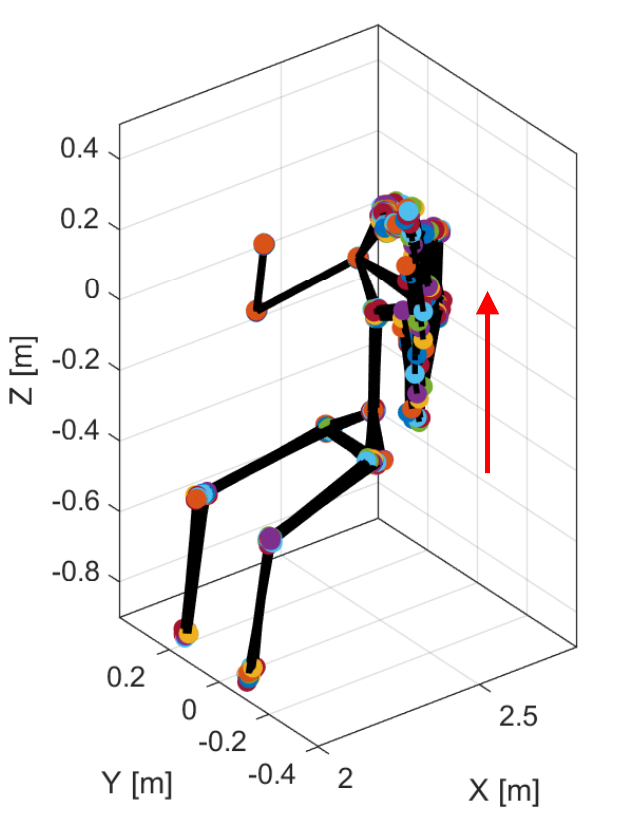}}
    \subfigure[]{\includegraphics[width=1.9in]{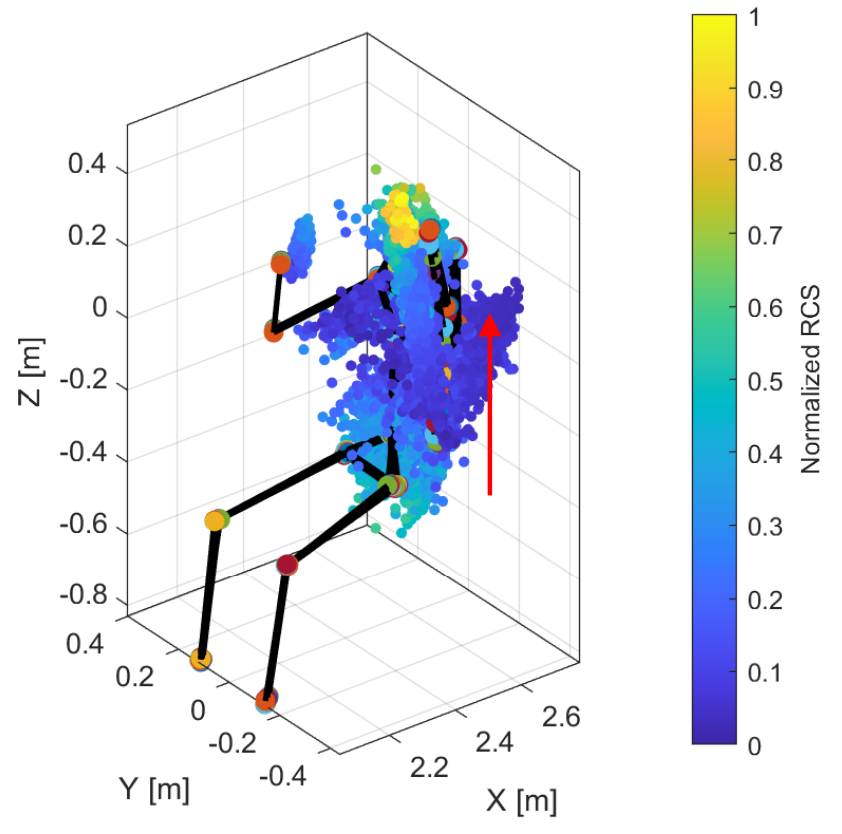}}
    \caption{The simulated scattering points on a new human gesture. (a) Aggregated human body keypoint positions during the gesture, showing motion trajectories. (b) Aggregated simulated scattering points.}
    \label{fig12}
\end{figure}

\subsection{Cross-Validation}

To evaluate the generalization of the proposed approach, we conduct two types of cross-validation experiments: (1) generation under new subjects, and (2) generation under new human gestures. In each case, the simulated scattering points are compared with the measured data in terms of spatial distribution and RCS over time.

Figure~\ref{fig11} shows the comparison results on a new subject performing a typical human gesture (\textit{Up-Up}). The generated scattering points closely match the measured ones in both spatial structure and RCS distribution. Across different human body parts, the average spatial error of the generated points is less than 5\,cm, which is computed as the average Euclidean distance between each generated scattering point and its corresponding matched measured point. Figure~\ref{fig12} presents the simulation results for a previously unseen gesture, where the right hand remains stationary while the left hand is raised. This gesture was not included in the training dataset, making it a good case for evaluating generalization performance. Despite the significant structural changes in 3D key-point models, the simulated scattering points still exhibit consistent regional distributions. These results indicate that the proposed method achieves good generalization performance.

\subsection{CIR Simulations}

\begin{figure}
\centering
\subfigure[]{\includegraphics[width=1.7in]{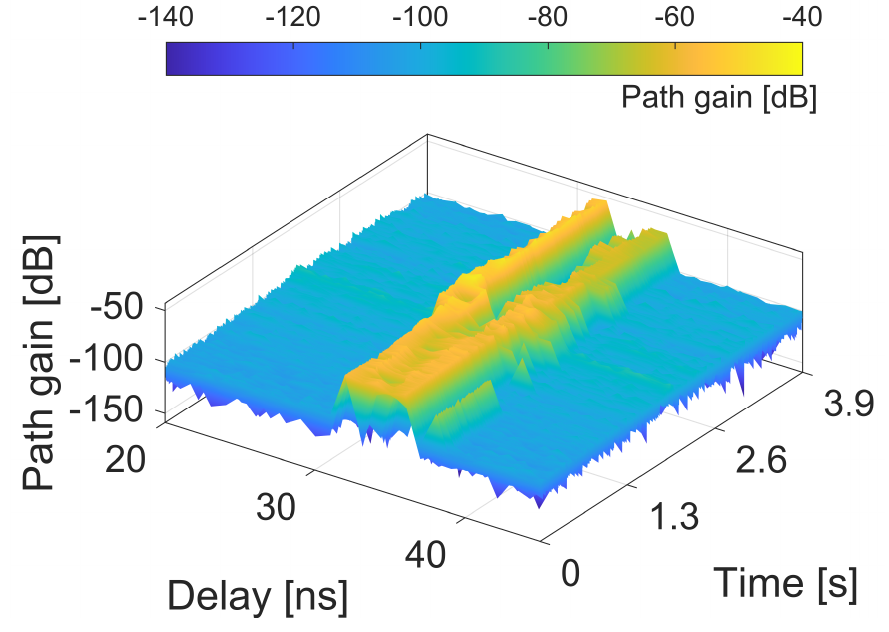}}
\subfigure[]{\includegraphics[width=1.7in]{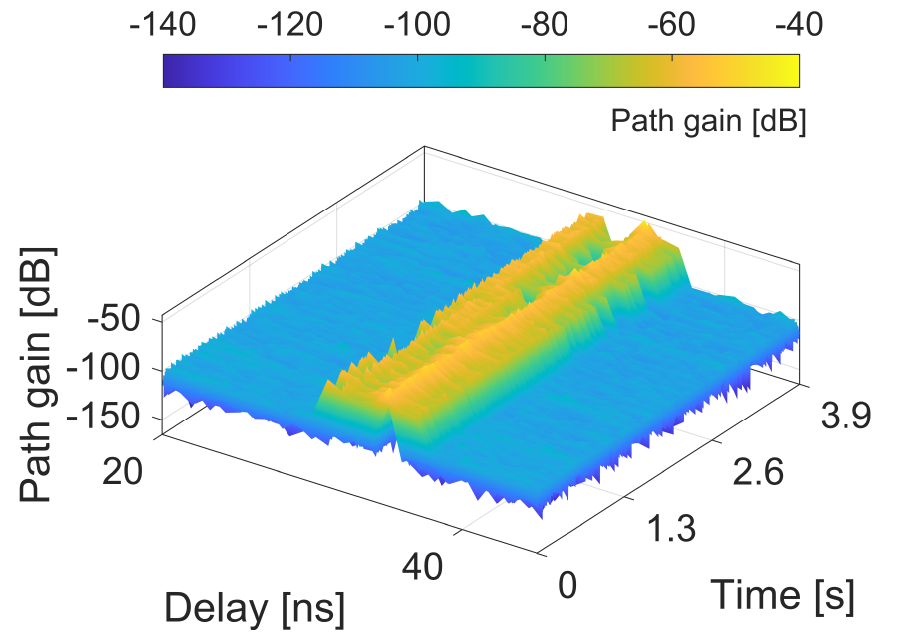}}
\caption{Comparison of measured and simulated PDP on gesture: \textit{Up-Up}. (a) Measured. (b) Simulated.}
\label{fig13}
\end{figure}

\begin{figure}
\centering
\subfigure[]{\includegraphics[width=1.7in]{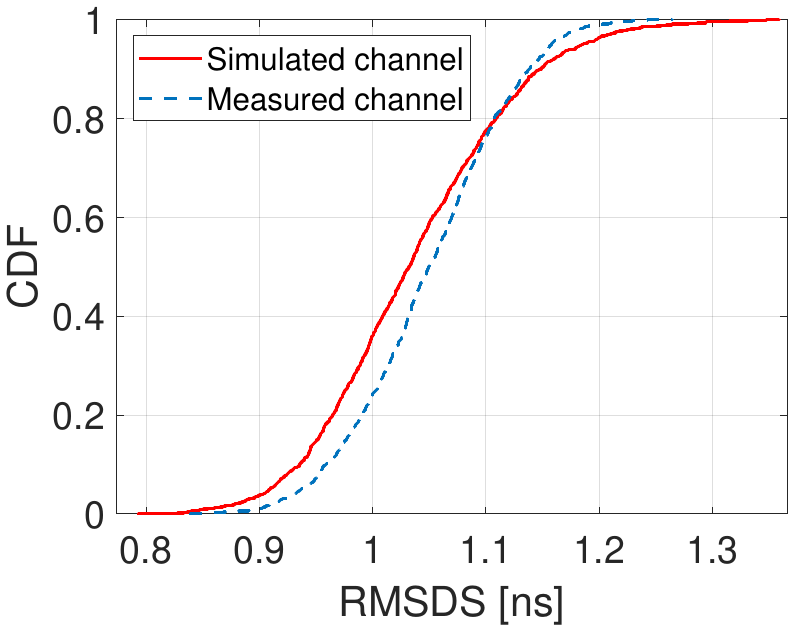}}
\subfigure[]{\includegraphics[width=1.7in]{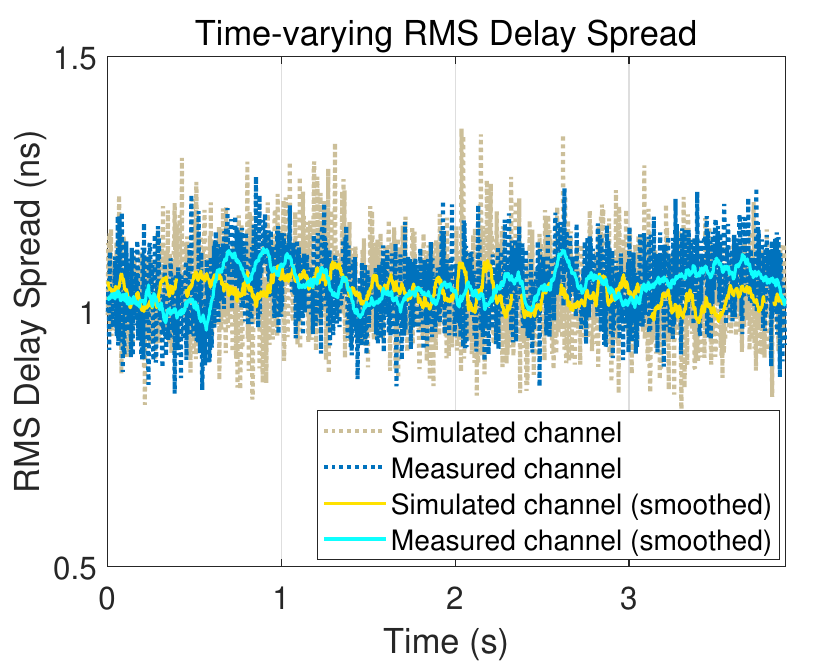}}
\caption{Comparison of measured and simulated RMS delay spread characteristics. (a) CDF of RMSDS. (b) Time-varying RMSDS.}
\label{fig14}
\end{figure}

To validate the accuracy of the proposed human gesture channel modeling in capturing propagation characteristics, we reconstruct the CIR based on the generated MPCs. Figure~\ref{fig13} compares the measured and simulated Power Delay Profiles (PDPs) under a typical \textit{Up-Up} gesture. It can be observed that the simulated PDP exhibits strong reflected paths that closely match the measured data, showing similar delay patterns and multipath distributions. This indicates that the generated channels are physically consistent in measured channels.

Furthermore, we analyze the statistical behavior of the Root Mean Square Delay Spread (RMSDS) between the measured and simulated CIRs. As shown in Figure~\ref{fig14}, the Cumulative Distribution Function (CDF) curves of RMSDS reveal highly consistent trends. The time-varying RMSDS curves also demonstrate similar fluctuation results, with the RMSDS error remaining within $0.1\ \mathrm{ns}$ for 90~\% of the snapshots. These results confirm that the proposed method not only accurately captures the statistical characteristics of the channel but also effectively reproduces temporal dynamics of the channel.

\subsection{Micro-doppler Signatures}

To further validate the proposed method in modeling micro-dynamic behaviors, this section analyzes the simulation results of micro-Doppler signatures, including the velocity variation trends of different body parts and the overall synthesized micro-Doppler signatures.

\begin{figure}[!ht]
    \centering
    \includegraphics[width=3.5in]{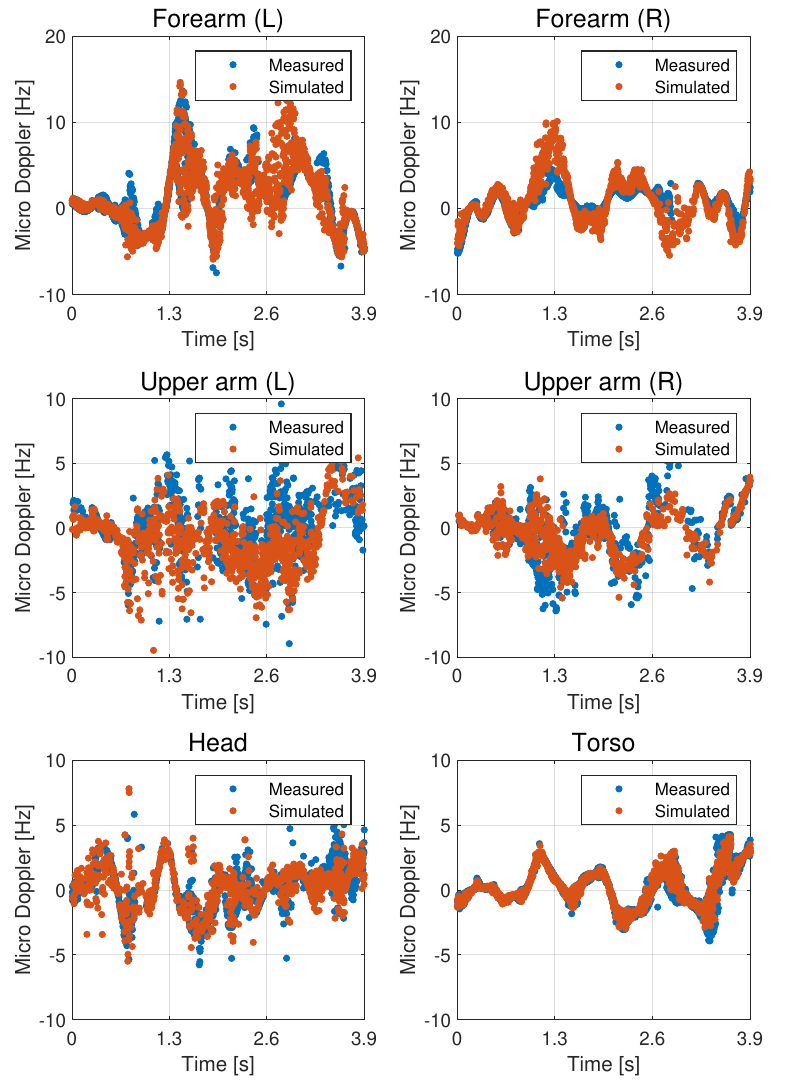} 
    \caption{Micro-doppler of different body parts on gesture: \textit{Up-Up}.}
    \label{fig15}
\end{figure}

\begin{figure}[!ht]
\centering
\subfigure[]{\includegraphics[width=1.65in]{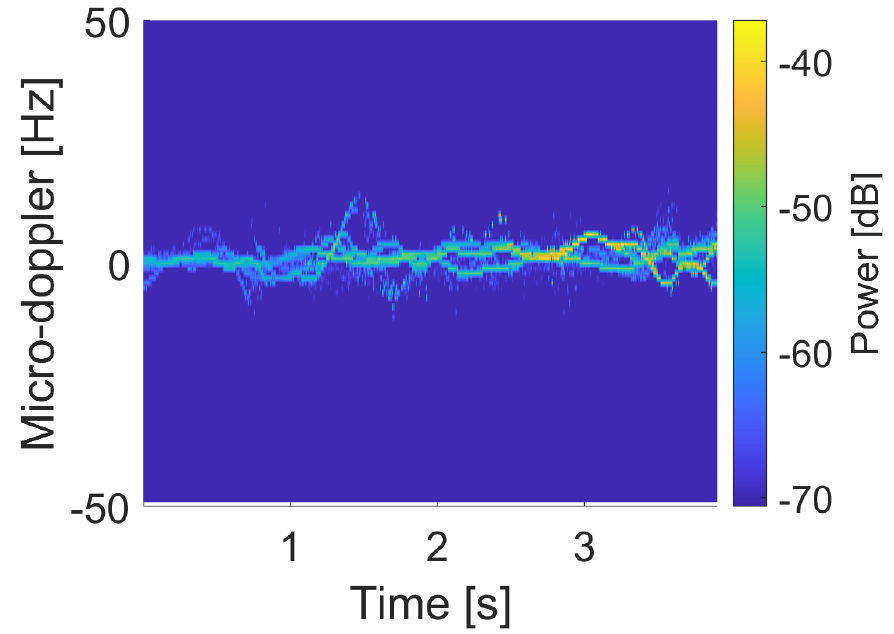}}
\subfigure[]{\includegraphics[width=1.65in]{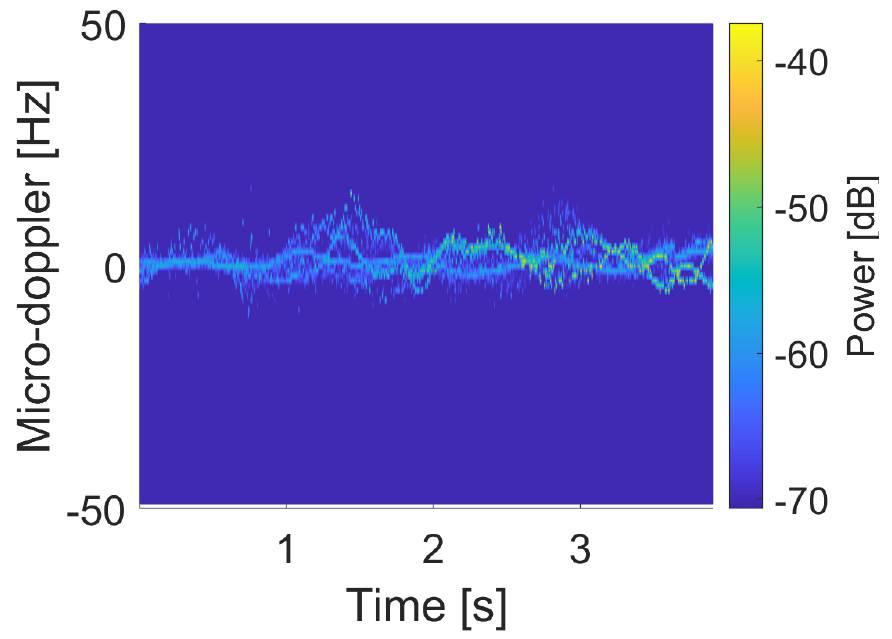}}
\subfigure[]{\includegraphics[width=1.65in]{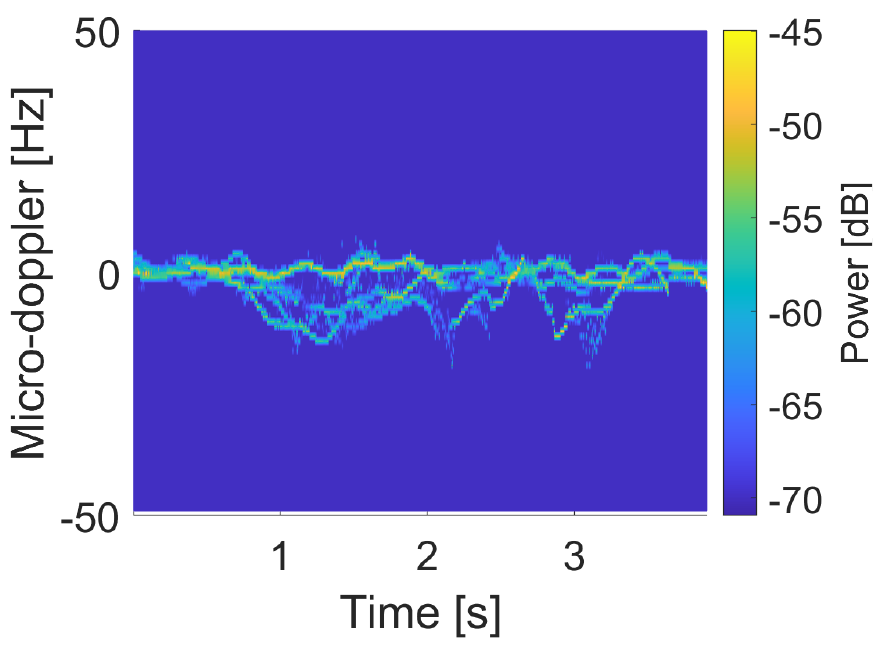}}
\subfigure[]{\includegraphics[width=1.65in]{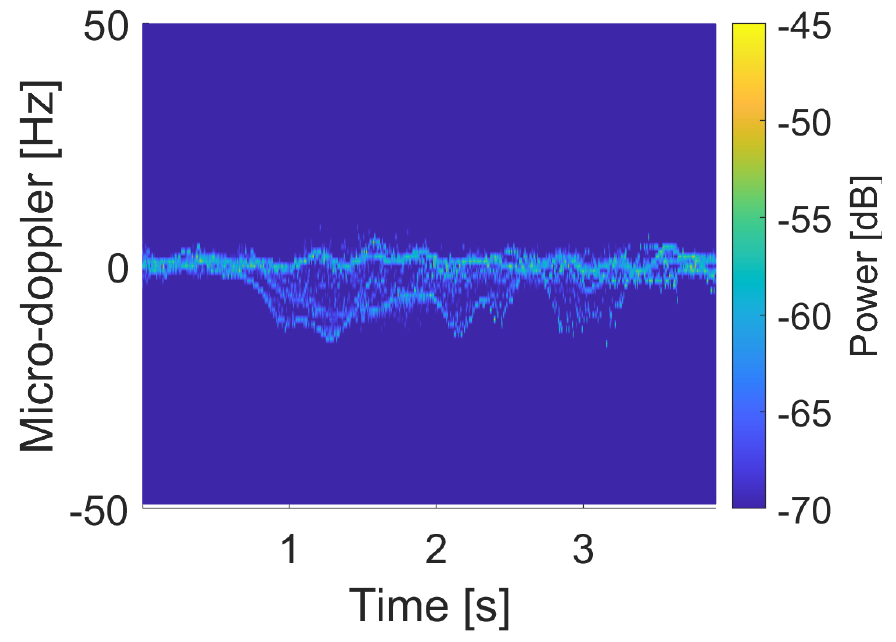}}
\subfigure[]{\includegraphics[width=1.65in]{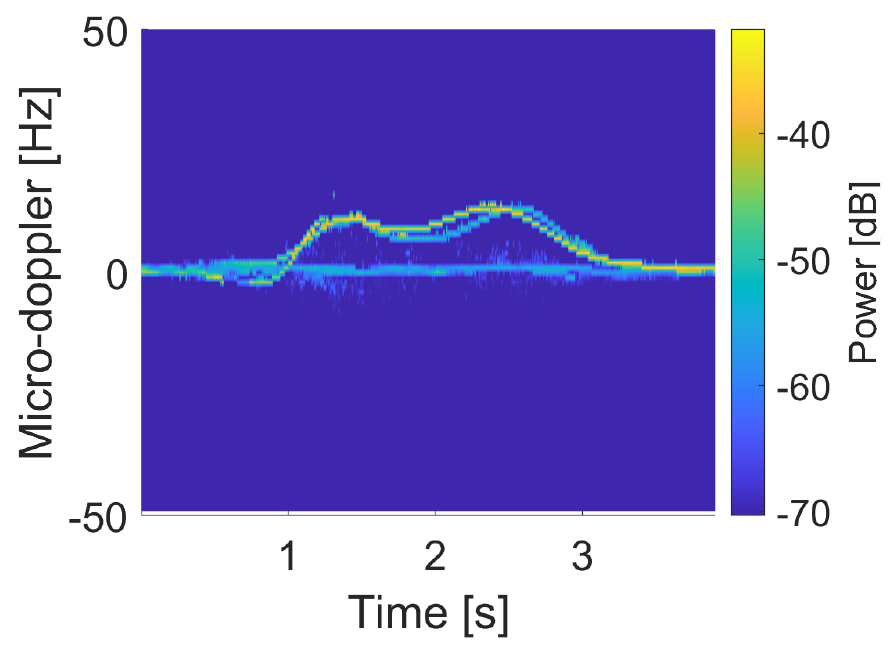}}
\subfigure[]{\includegraphics[width=1.65in]{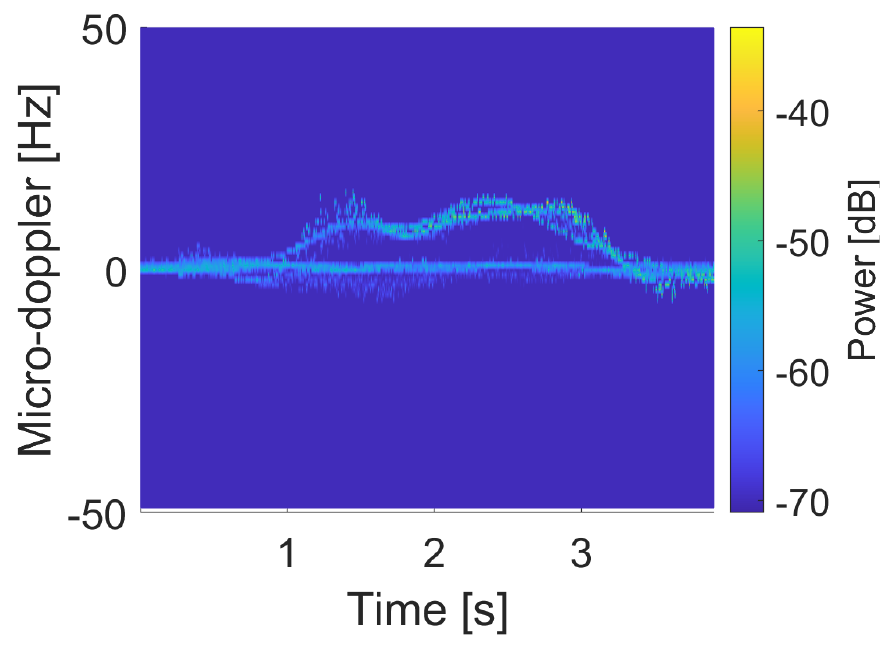}}
\subfigure[]{\includegraphics[width=1.65in]{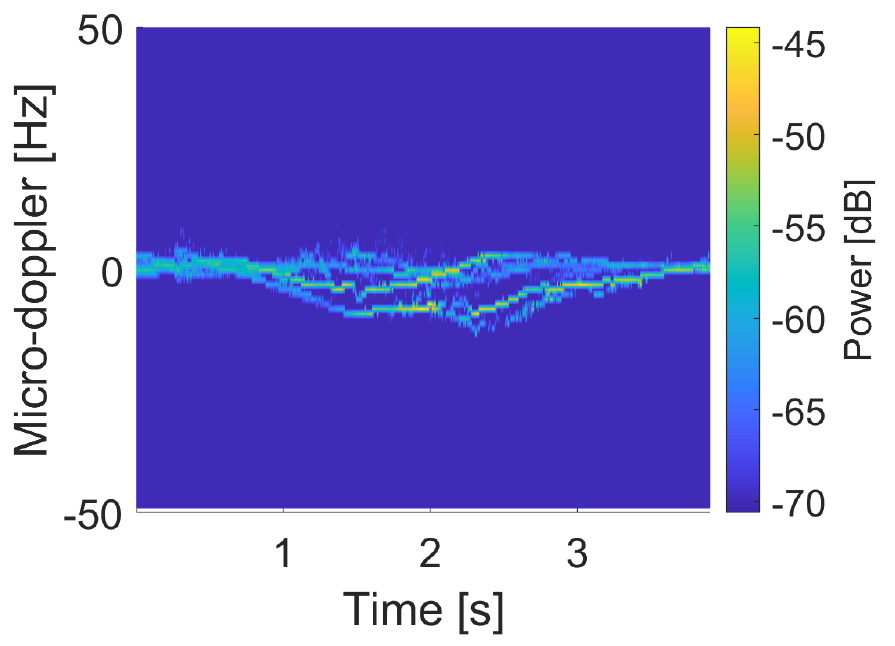}}
\subfigure[]{\includegraphics[width=1.65in]{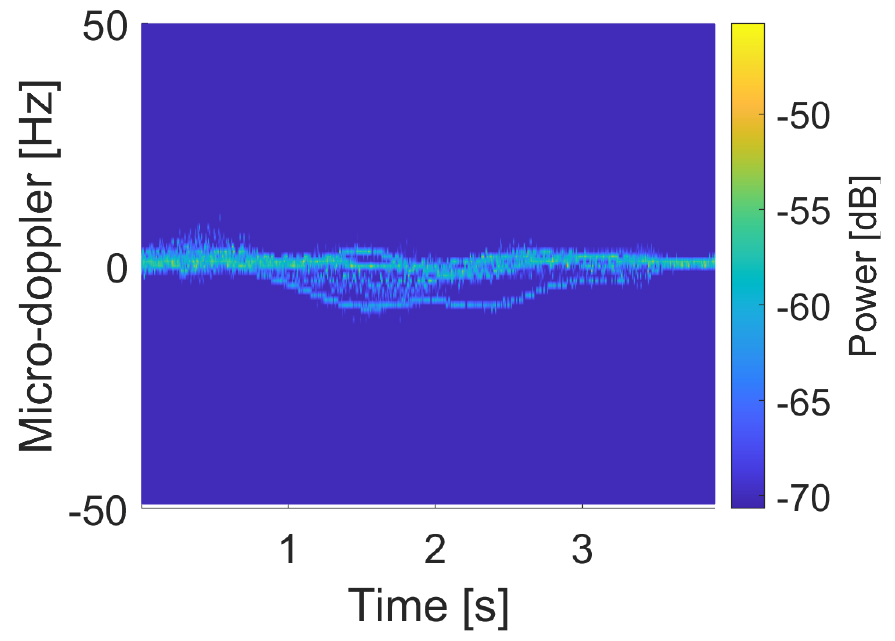}}
\caption{Comparison of measured and synthetic micro-Doppler signatures. 
(a) Measured micro-Doppler signature of the \textit{Up–Up} gesture; 
(b) Synthetic counterpart of (a); 
(c) Measured micro-Doppler signature of the \textit{Right–Left} gesture; 
(d) Synthetic counterpart of (c); 
(e)-(f) Measured and synthetic micro-Doppler signatures of the \textit{Up–Up} gesture performed by a different subject;
(g)-(h) Measured and synthetic micro-Doppler signatures of the \textit{Right–Left} gesture performed by a different subject.}
\label{fig16}
\end{figure}

Figure~\ref{fig15} presents the micro-Doppler trajectories of six human body parts under the \textit{\textit{Up-Up}} gesture. It can be observed that the simulated Doppler characteristics exhibit a high degree of consistency with the measured data in terms of variation trends, amplitude range, and the location of peaks. This confirms that the proposed model can accurately capture the impact of local body motion on micro-Doppler effects.

Furthermore, Figure~\ref{fig16} provides the comparison results of overall micro-Doppler signatures under different gestures and subjects. Figure~\ref{fig16} (a)--(d) illustrate the comparisons between the real and synthetic micro-Doppler signatures for \textit{\textit{Up-Up}} and \textit{Right-Left} gestures. The results show that the proposed model can effectively reproduce the micro-Doppler trajectory patterns caused by different gestures. Figure~\ref{fig16} (e)--(h) further verify the generalization capability across different subjects under the same gesture \textit{\textit{Up-Up}} and \textit{Right-Left}. The synthesized micro-Doppler signatures closely match the measured ones in terms of energy distributions and motion trajectories, demonstrating the ability to reproduce micro-Doppler shifts of human body variations.

\section{Conclusion}

In this paper, we propose a deep learning-based human gesture channel modeling framework for ISAC scenarios. The method decomposes the human body into different human body parts and models the multipath propagation characteristics within each region using deep neural networks. Context-aware millimeter-wave sensing channel measurements are employed to obtain channel responses and human gesture information. Based on actual measurements, a Poisson neural network is trained to predict MPC number and C-VAEs are reused to generate scattering points under different gesture conditions. The generated scattering points are further used to reconstruct temporally continuous CIRs and micro-Doppler signatures. The simulation results demonstrate that the proposed method achieves good accuracy in capturing channel characteristics and exhibits strong generalization across different subjects and gestures. Furthermore, the framework offers an interpretable approach for human gesture channel simulation, which can be used for data augmentation and the evaluation of gesture-based ISAC systems.


\bibliographystyle{IEEEtran}

\bibliography{refs}

\newpage

\vspace{11pt}

\vspace{11pt}

\vfill

\end{document}